\documentclass[useAMS,usenatbib]{mn2e}

\usepackage{amsmath}
\usepackage{graphicx}
\usepackage{amssymb}

\newcommand{\xij}{\xi_{ij}^{-1}}
\newcommand{\xik}{\xi_{ik}^{-1}}
\newcommand{\xil}{\xi_{il}^{-1}}
\newcommand{\xjk}{\xi_{jk}^{-1}}
\newcommand{\xjl}{\xi_{jl}^{-1}}
\newcommand{\xkl}{\xi_{kl}^{-1}}

\newcommand{\nl}{\mathrm{NL}}
\newcommand{\nside}{\textsc{\small{NSIDE}}}

\title[f$_\nl$ and g$_\nl$ constraints from the CMB]
{Constraints on f$_\nl$ and g$_\nl$ from the analysis of the N-pdf of the CMB large scale anisotropies}
\author[P. Vielva and J. L. Sanz]
{P. Vielva$^1$, J. L. Sanz$^{1}$ \\
$^{1}$Instituto de F{\'\i}sica de Cantabria (CSIC - Univ. de Cantabria), Avda. Los Castros s/n, 39005 - Santander, Spain\\
\hspace{0.1cm}E-mails : vielva@ifca.unican.es, sanz@ifca.unican.es
\\}
\date{Accepted ???. Received ???; in original form \today}
\pagerange{\pageref{firstpage}--\pageref{lastpage}}
\pubyear{1998}

\begin{document}
\maketitle
\label{firstpage}

\begin{abstract}
In this paper we extend a previous work~\citep{vielva09} where we presented a method based on the N-point probability
distribution (pdf) to study the Gaussianity of the cosmic microwave background (CMB).
We explore a local
non-linear perturbative model up to third order as a general characterization of the CMB anisotropies.
We focus our analysis in large scale anisotropies ($\theta > 1^\circ$).
At these angular scales (the Sachs-Wolfe regime), the
non-Gaussian description proposed in this work defaults (under certain
conditions) to an approximated local form of the weak non-linear
coupling inflationary model. In particular, the \emph{quadratic} and \emph{cubic} terms are
governed by the non-linear coupling parameters f$_\nl$ and g$_\nl$, respectively.
The extension proposed in this paper allows us to directly constrain these non-linear parameters.
Applying the proposed methodology to WMAP 5-yr data, we obtain 
$-5.6\times10^5 < {\rm g}_{\nl} < 6.4\times10^5$, at 95\% CL. This result is in agreement with
previous findings obtained for equivalent non-Gaussian models and
with different non-Gaussian estimators, although this is the first direct constrain on g$_\nl$ from CMB data.
A model selection test is performed, indicating that a Gaussian model 
(i.e. f$_\nl \equiv 0$ and g$_\nl \equiv 0$  ) is preferred to the non-Gaussian scenario.
When comparing different non-Gaussian models, we observe that a pure f$_\nl$ model (i.e. g$_\nl \equiv 0$)
is the most favoured case, and that a pure g$_\nl$ model (i.e. f$_\nl \equiv 0$) is more likely than
a general non-Gaussian scenario (i.e. f$_\nl \neq 0$ and g$_\nl \neq 0$).
Finally, we have analyzed the WMAP data in two independent hemispheres, in particular the ones defined by the
dipolar pattern found by~\cite{hoftuft09}. We show that, whereas the g$_\nl$ value is compatible between both
hemispheres, it is not the case for f$_\nl$ (with a p-value $\approx 0.04$). However, if, as suggested 
by~\cite{hoftuft09}, anisotropy of the data is assumed, the distance between the likelihood
distributions for each hemisphere is larger than expected from Gaussian and anisotropic simulations,
not only for f$_\nl$, but also for g$_nl$ (with a p-value of $\approx 0.001$ in the case of this latter
parameter).
This result is an extra evidence for the CMB asymmetries previously reported in WMAP data.
\end{abstract}


\section{Introduction}
\label{intro}

Current observations of the Cosmic Microwave Background (CMB) temperature and polarization
fluctuations, in addition to other astronomical data sets~\citep[e.g.][see \citealt{barreiro09} for a recent review]{komatsu09,gupta09}, 
provide an overall picture for the origin,
evolution, and matter and energy content of the universe, which is usually referred to as the
\textit{standard cosmological model}. In this context, we believe the universe to be
highly homogeneous and isotropic, in expansion, well 
described by a Friedmman-Robertson-Walker metric and with a trivial topology.
The space geometry is very close to flat, and it is filled with cold dark matter (CDM) and dark energy 
(in the form of a cosmological constant, $\Lambda$), in addition to baryonic matter and electromagnetic 
radiation. Large scale structure (LSS) is assumed to be formed by the gravitational collapse of 
an initially smooth distribution of adiabatic matter fluctuations, which were seeded by initial Gaussian 
quantum fluctuations generated in a very early inflationary stage of the universe evolution.

It is interesting to mention that, besides the success of current high quality CMB 
data~\citep[in particular the data provided by the WMAP satellite][]{hinshaw09} in
constraining the cosmological parameters with good accuracy and in
showing the high degree of homogeneity and isotropy of the 
Universe~\citep[as predicted by the standard inflation scenario, see for instance][]{liddle00},
it has been, precisely, through the analysis of this very same data, that the
CMB community has been allowed to probe fundamental principles and assumptions of the
\emph{standard cosmological model}. Most notably, the application of
sophisticated statistical analysis to CMB data might help us to
understand whether the temperature fluctuations of the primordial
radiation are compatible with the fundamental isotropic and Gaussian
standard predictions from the \emph{inflationary phase}. 

Indeed, the interest of the cosmology community in this field has experienced a
significant growth, since several analyses of the WMAP data have reported
some hints for departure from isotropy and Gaussianity of the CMB
temperature distribution. The literature on the subject is very large, and
is still growing, which makes really difficult to provide a complete and updated
list of publications. We refer to our previous work~\citep{vielva09} for an (almost) complete
review.

Among the previously mentioned analyses, those related to the study of
non-standard inflationary models have attracted a
larger attention. For instance, the
non-linear coupling parameter f$_{\nl}$ that describes the
non-linear evolution of the inflationary potential \citep[see
e.g.][and references therein]{bartolo04} has been constrained by
several groups, from the analysis of the WMAP data: using the angular bispectrum
\citep{komatsu03,creminelli07,spergel07,komatsu09,yadav08,smith09}; applying the
Minkowski functionals
\citep{komatsu03,spergel07,gott07,hikage08,komatsu09}; using
different statistics based on wavelets
\citep{mukherjee04,cabella05,curto09a,curto09b,pietrobon09,rudjord09}, and
by exploring the N-pdf
\citep{vielva09}. Besides marignal detections of f$_{\nl} > 0$~\citep[with a probability of around 95\%,][]{yadav08,rudjord09},
there is a
general consensus on the WMAP compatibility with the predictions
made by the standard inflationary scenario at least at 95\%
confidence level. The current best limits obtained from the CMB data are:
$-4 < {\rm f}_{\nl} < 80$ at 95\% CL by~\cite{smith09}. In addition, very recently, promising constraints
coming out from the analysis of LSS have been reported: $-29 < {\rm f}_{\nl} < 70$ at 95\% CL~\citep{slozar08}.

The aim of this paper is to extend our previous work~\citep{vielva09}, 
where the full N-pdf of a non-Gaussian model  that describes
the CMB anisotropies as a local (pixel-by-pixel) non-linear expansion of the temperature fluctuations (up to second order) was derived.
For this model ---that, at large scales, can be considered
as an approximation to the weak non-linear coupling
scenario--- we are able to build the exact likelihood on
pixel space. Working in pixel space allows one to include easily
non-ideal observational conditions, like incomplete sky coverage and
anisotropic noise. The extension of the present work is to account for higher order moments
in the expansion, in particular, we are able to directly obtain constraints on g$_\nl$, that is
the coupling parameter governing the cubic term of the weak non-linear expansion.

As far as we are aware, direct constraints on g$_\nl$ have been made available only very recently \citep{desjacques09}
and are obtained from LSS analyses: $-3.5\times 10^{5} < {\rm g}_{\nl} < 8.2\times 10^{5}$ at 95\% CL.
This constraint was obtained for the specific case in which the coupling parameter governing the quadratic
term (f$_\nl$) is negligible~\citep[i.e. f$_\nl \equiv 0$, which is the situation required for some 
curvaton inflationary models, e.g.][]{sasaki06,enqvist08,huang09}.
We present in this work the first direct measurement of g$_\nl$ obtained from CMB data. In addition to
study the particular case of f$_\nl \equiv 0$, we also consider a more general case in which a joint estimation
of f$_\nl$ and g$_\nl$ is performed.
Finally, and justified by recent findings~\citep[e.g.,][]{hoftuft09}, we compute the N-pdf in two different hemispheres, and
derive from it constraints on f$_\nl$ and g$_\nl$ for this hemispherical division of the celestial sphere.

The paper is organized as follows. In Section~\ref{sec:model} we
describe the physical model based on the local expansion of the CMB
fluctuations and derive the full posterior probability, recalling how it defaults to
the case already addressed in \cite{vielva09}. In Section~\ref{sec:simulations}
we check the methodology against WMAP-like simulations. Results on 
WMAP 5-year data are presented in Section~\ref{sec:wmap}. Conclusions are given
in Section~\ref{sec:final}. Finally, in Appendix~\ref{sec:app}, we provide a detailed computation of the full N-pdf.

\section{The non-Gaussian model}
\label{sec:model}

Although current CMB measurements are well described by random Gaussian anisotropies
(as it is predicted by the standard inflationary scenario), observations also
allows for small departures from Gaussianity, that could indicate the presence
of an underlying physical process generated in non-standard models.

As we did in \cite{vielva09} we adopt a parametric non-Gaussian model for the CMB anisotropies,
that accounts for a small and local (i.e. point-to-point)
perturbation of the CMB temperature fluctuations, around its
intrinsic Gaussian distribution:
\begin{eqnarray}
\label{eq:physical_model} 
{\Delta T}_i & = &  \left({\Delta T}_i\right)_G + a\left[\left({\Delta T}_i\right)_G^2 - \left\langle\left({\Delta T}_i\right)_G^2\right\rangle \right] + \nonumber \\ 
& &  b\left({\Delta T}_i\right)_G^3 + {\cal O}\left(\left({\Delta T}_i\right)_G^4\right).
\end{eqnarray}
The \emph{linear term} ($\left({\Delta T}_i\right)_G$) 
is given by a Gaussian N-point probability density function (N-pdf) that is
easily described in terms of the standard inflationary model. The
second and third terms on the right-hand side are the \emph{quadratic} and the
\emph{cubic} perturbation terms, respectively, and they are governed by the
$a$ and $b$ parameters. 

The sub-index
$i$ refers to a given direction in the sky that, in practice, is described
in terms of a certain pixelization of the sphere. The operator $\langle \cdot \rangle$ 
indicates averaging over all the pixels defining the sky coverage.
We have not considered explicitly an \emph{instrumental noise}-like term, since, for the particular
case that we intend to explore (i.e., large-scale CMB data), its contribution to the
measured signal (for experiments like WMAP or Planck) is negligible. Precisely at the large-scale regime, the term
${\Delta T}_i$ is mostly dominated by the Sachs-Wolfe contribution to the CMB fluctuations,
and can be related to the primordial gravitational potential $\Phi$~\citep[e.g.][]{komatsu01} by:
\begin{equation}
\label{eq:sw_limit} {\Delta T}_i \approx -\frac{T_0}{3} \Phi_i,
\end{equation}
where $T_0$ is the CMB temperature. Small departures from Gaussianity of the $\Phi$ potential are usually described via the weak 
non-linear coupling model:
\begin{equation}
\label{eq:weak_model} \Phi_i = {\Phi_L}_{,i} + 
{\rm f}_\nl\left( {\Phi_L}_{,i}^2 - \langle {\Phi_L}_{,i}^2 \rangle \right) + 
{\rm g_\nl} {\Phi_L}_{,i}^3 + {\cal O}\left( {\Phi_L}_{,i}^4 \right).
\end{equation}
Taking into account equations~\ref{eq:physical_model},~\ref{eq:sw_limit} and~\ref{eq:weak_model}, and always
considering the specific case for scales larger than the
horizon scale at the recombination time (i.e. above the degree
scale), it is trivial to establish the following
relations:
\begin{equation}
\label{eq:relation} {\rm f}_\nl \cong -\frac{T_0}{3}a ~{\rm ,}~~~~~~~ {\rm g}_\nl \cong \left(\frac{T_0}{3}\right)^2b.
\end{equation}
At this point, it is worth mentioning that the model in equation~\ref{eq:physical_model} does not
pretend to incorporate all the
gravitational and non-gravitational effects, due to the evolution of the
initial quadratic potential model, but rather allows for a
useful parametrization for describing a small departure from Gaussianity. The relationships
in equation~\ref{eq:relation} have to be understood just as an asymptotic equivalence for large scales.

Let us simplify the notation by
transforming the Gaussian
term $\left({\Delta T}_i\right)_G$ into
a zero mean and unit variance random variable $\phi_i$. It is easy to show that equation~\ref{eq:physical_model} 
transforms into:
\begin{equation}
\label{eq:model} x_i = \phi_i + a\sigma\left(\phi_i^2 - 1\right) +
b\sigma^2\phi_i^3+ {\cal O}\left(\sigma^3\right)
\end{equation}
where:
\begin{equation}
\label{eq:equivalences} x \equiv \frac{1}{\sigma}{\Delta T} ~{\rm ,}~~~~~~~ \phi \equiv \frac{1}{\sigma}\left({\Delta T}\right)_G 
\end{equation}
and $\sigma^2 \equiv \left\langle\left({\Delta
T}_i\right)_G^2\right\rangle$ is the standard deviation of the CMB fluctuations.
Trivially, the normalized Gaussian variable $\phi$ satisfies: 
\begin{eqnarray}
\label{eq:phiproperties} 
\langle \phi_i^{2n+1} \rangle & = & 0 \nonumber \\
\langle \phi_i^{2m} \rangle & = & \left(2m-1\right)!! \nonumber \\
\langle \phi_i \phi_j \rangle & = & \xi_{ij},
\end{eqnarray}
where $n \geq 0$ and $m > 0$ are integer numbers, and $\xi_{ij}$ represents the normalized correlation between 
pixels $i$ and $j$. Obviously, the N-pdf of the $\bmath{\phi}=\{\phi_1, \phi_2, ...,
\phi_N\}$ random field (where $N$ refers to the number of pixels on
the sphere that are actually observed) is given by a multivariate Gaussian distribution:
\begin{equation}
\label{eq:pdfphi} p(\bmath{\phi}) = \frac {1}{(2\pi
)^{N/2}(\det{\bmath{\xi}} )^{1/2}}e^{-\frac{1}{2}
\bmath{\phi}\bmath{\xi}^{-1}\bmath{\phi}^{t}},
\end{equation}
where $\bmath{\xi}$ denotes the correlation matrix and operator
$\cdot^{t}$ denotes standard matrix/vector transpose.

As it was the case in~\cite{vielva09}, the objective is to obtain the N-pdf 
associated to the non-Gaussian $\bmath{x}=\{x_1, x_2, ..., x_N\}$ field, as a function of the
non-linear coupling parameters (or, equivalently, the $a$ and $b$ coefficients):
\begin{equation}
\label{eq:pdfx} p(\bmath{x}\vert a,b ) = p(\bmath{\phi} =
\bmath{\phi} (\bmath{x})) Z.
\end{equation}
In this expression, $Z$ is the determinant of the Jacobian for the $\bmath{\phi} \longrightarrow \bmath{x}$
transformation. Because the proposed model is local (i.e. point-to-point), the Jacobian matrix
is diagonal and, therefore, $Z$ is given by:
\begin{equation}
\label{eq:jacobian} Z = \det{\left[ \frac{\partial \phi_i}{\partial
x_j} \right]} = \prod_i \left( \frac{\partial \phi_i}{\partial
x_i}\right).
\end{equation}

Both, equations~\ref{eq:pdfx} and~\ref{eq:jacobian}, require the inversion of equation~\ref{eq:model}:
i.e., to express $\phi_i$ as a function of $x_i$. After some algebra, it can be proved that:
\begin{equation}
\label{eq:inversmodel}
\phi_i = x_i + \eta_i \sigma + \nu_i \sigma^2 + \lambda_i \sigma^3 + \mu_i \sigma^4 + {\cal O}\left(\sigma^5\right),
\end{equation}
where:
\begin{eqnarray}
\label{eq:numu}
\eta_i & = & -a\left(x_i^2 - 1\right) \nonumber \\
\nu_i & = & \left(  2a^2 - b \right)   x_i^3 - 2a^2x_i \nonumber \\
\lambda_i & = & \left( 5ab - 5a^3 \right)  x_i^4 + \left( 6a^3 - 3ab \right)  x_i^2  - a^3 \nonumber \\
\mu_i & = & \left( 14a^4 - 21a^2b + 3b^2 \right)  x_i^5 + \left(-20a^4 +20a^2b\right)  x_i^3 \nonumber \\
& &  + \left( 6a^4 - 3a^2b \right)  x_ i.
\end{eqnarray}
Instead of dealing with $p(\bmath{x}\vert a,b )$, it is equivalent, but more convenient, to work with the 
log-likelihood ${\mathcal L} \left(\bmath{x}|a,b\right)$:
\begin{equation}
\label{eq:loglike}
{\mathcal L} \left(\bmath{x}|a,b\right) = \log \frac{p\left(\bmath{x}\vert a,b \right)}{p\left(\bmath{x}\vert 0 \right)}.
\end{equation}
A detailed computation of ${\mathcal L} \left(\bmath{x}|a,b\right)$ is given in Appendix A. Let us just recall here
its final expression:
\begin{eqnarray}
\label{eq:loglikegeneral} 
\frac{1}{N} {\mathcal L}  \left(\bmath{x}|a,b\right) & = & F\sigma + \left(2a^2 - 3b + G\right)\sigma^2  + H\sigma^3 \nonumber \\
& & + \left(12a^4 - 36a^2b + \frac{27}{2}b^2 + I \right)\sigma^4,
\end{eqnarray}
where $N$ is the number of data points, and $F$, $G$, $H$ and $I$ are 
functions of $a$ and $b$ (see~\ref{eq:coeffs}). The desired N-pdf, $p(\bmath{x}\vert a,b)$ is 
obtained by the inversion of equation~\ref{eq:loglike}, and taking into account that 
$p\left(\bmath{x}\vert 0 \right) \equiv p\left(\phi = x\right)$, i.e., the known Gaussian N-pdf in equation~\ref{eq:pdfphi}.

\section{Application to WMAP simulations}
\label{sec:simulations}

In this Section we aim to investigate the performance of the parameters
estimation from the N-pdf derived in the previous Section. We explore
different non-Gaussian scenarios; in particular, we 
study three particular cases of special interest:
\begin{itemize}
\item \emph{Case i}) $a\neq0$, $b=0$. This scenario would correspond, for example, to the case for the slow-roll standard inflation

\item \emph{Case ii}) $a=0$, $b\neq0$. This scenario would correspond to the particular situation for some curvaton models.

\item \emph{Case iii}) $a\neq0$, $b\neq0$. It is a generic case, not representing any specific inflationary
model, but rather a general scenario.
\end{itemize}

In particular, we will study how the determination of the parameters governing the
non-Gaussian terms is performed, and what is the
impact when one is exploring different configurations. In the next subsections, we will focus, first, in
the case in which a slow-roll standard like scenario is assumed (i.e., we only try to adjust for the \emph{quadratic} term,
assuming the \emph{cubic} one is negligible), whatever the data is actually a pure \emph{quadratic} or \emph{cubic} model, or a
general non-Gaussian scenario. Second, we will follow a similar analysis, but assuming the estimation of a pure \emph{cubic} term. 
Finally, we
will address the case for a joint estimation of both (\emph{quadratic} and \emph{cubic}) terms. In the following, we will refer 
all our results in terms of the non-linear coupling parameters (f$_\nl$ and g$_\nl$), rather than to the $a$ and $b$ coefficients.

In order to carry out this analysis, we have generated Gaussian CMB simulations coherent with the model induced from the WMAP
5-year data at \nside=32 HEALPix~\citep{gorski05} resolution ($\approx 2^\circ$).

The procedure to generate a CMB Gaussian simulation ---$\left(\Delta
T\right)_G$ in equation~\ref{eq:physical_model}--- is as follows.
First, we simulate WMAP observations for the Q1, Q2, V1,
V2, W1, W2, W3, W4 difference assemblies at \nside=512 HEALPix
resolution. The $C_\ell$ obtained with the cosmological parameters
provided by the best-fit to WMAP data alone~\citep[Table 6
in][]{hinshaw09}, are assumed.

Second, a single co-added CMB map is computed afterwards through a
noise-weighted linear combination of the eight maps (from Q1 to W4).
The weights used in this linear combination are proportional to the inverse mean noise variance
provided by the WMAP team. They are
independent on the position (i.e., they are uniform
across the sky for a given difference assembly) and they are
forced to be normalized to unity. 
Notice that we have not added Gaussian white noise to the different difference assembly maps,
since we have already checked that instrumental noise plays a
negligible role at the angular resolution in which we are interested~\citep[$\approx 2^\circ$, see][for details]{vielva09}.

\begin{figure}
\includegraphics[width=8cm,keepaspectratio]{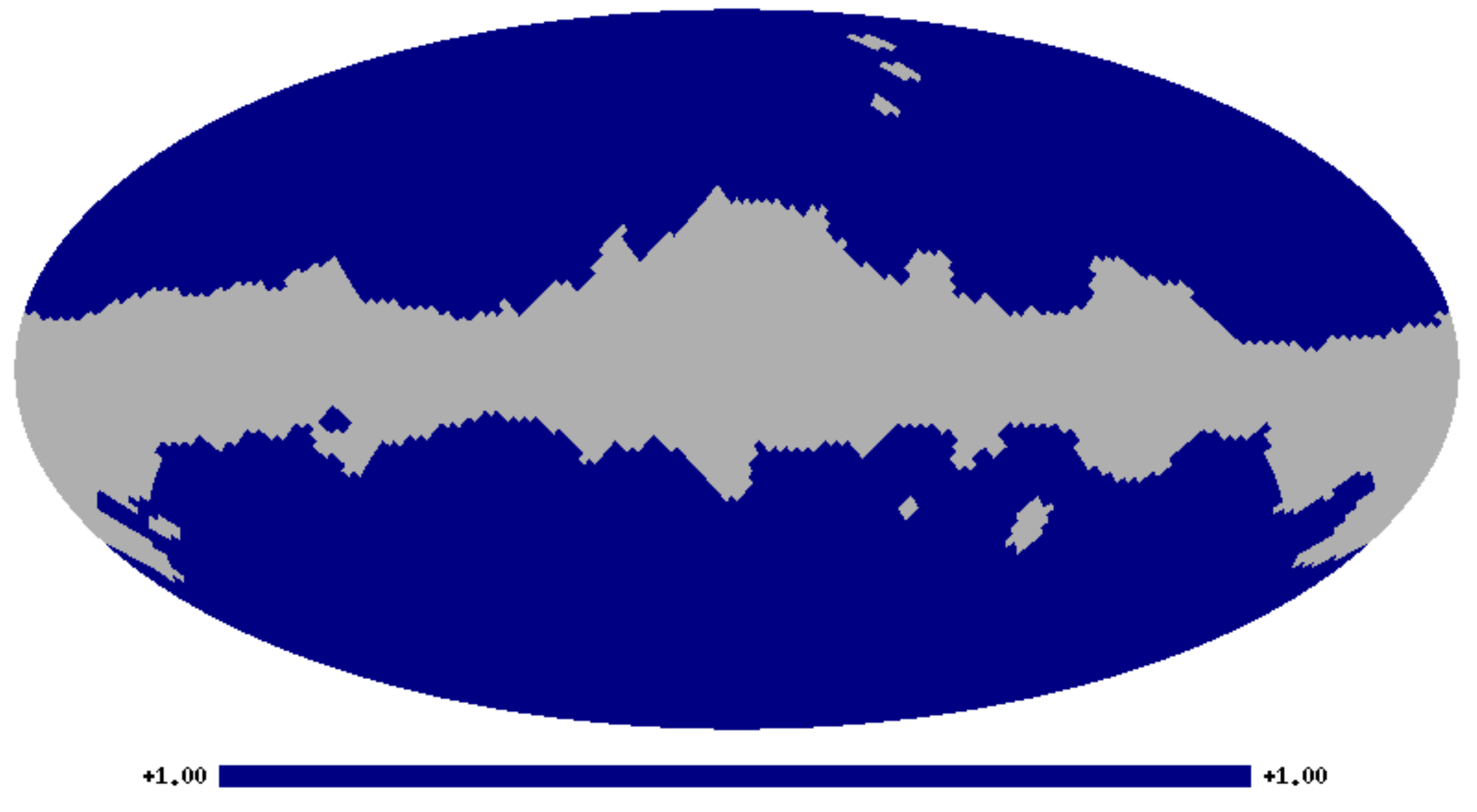}
\caption{\label{fig:mask}Mask at \nside=32 HEALPix resolution used
in this work. It corresponds to the WMAP KQ75 mask, although the
point source masking has not been considered, since the point
like-emission due to extragalactic sources is negligible at the
considered resolution. At this pixel resolution, the mask keeps around
69\% of the sky.}
\end{figure}

Third, the co-added map at
\nside=512 is degraded down to the final resolution of \nside=32, and
a mask representing a sky coverage like the one allowed by
the WMAP KQ75 mask~\citep{gold09} is adopted. At \nside=32 the
mask keeps around 69\% of the sky.
The mask is given in figure~\ref{fig:mask}. Let us remark that observational constraints
from an incomplete sky coverage can be easily taken into account by the
local non-Gaussian model proposed in this work, since it is
naturally defined in pixel space. This is not the case for other common estimators like
the bispectrum, where the presence of an incomplete sky coverage is usually translated into a
loss of efficiency.

We have generated 500000 simulations of $\left(\Delta T\right)_G$, computed as
described above, to estimate the correlation matrix $\bmath{\xi}$
accounting for the Gaussian CMB cross-correlations. We have
checked that this large number of simulations is enough to obtain an
accurate description of the CMB Gaussian fluctuations.

In addition, we have generated another set of 1000 simulations. These are
required to carry out the statistical analysis to check the performance
of the parameter estimation.
Each one of these 1000 $\left(\Delta T\right)_G$ simulations are
transformed into $\bmath{x}$ (following
equations~\ref{eq:physical_model} and~\ref{eq:equivalences}) to
study the response of the statistical tools as a function of the
non-linear parameters defining the local non-Gaussian
model proposed in equation~\ref{eq:model}.

Finally, let us remark that hereinafter the likelihood maximization is simply performed
by exploring a grid of values in the parameter space of the non-linear coupling
parameters. The step used in the grid is small enough to guarantee a good estimation
both of the likelihood peak and tails.

\subsection{The recovery of f$_\nl$ in the presence of a \emph{cubic} term}

\begin{figure*}
\begin{center}
\includegraphics[angle=90,width=15cm]{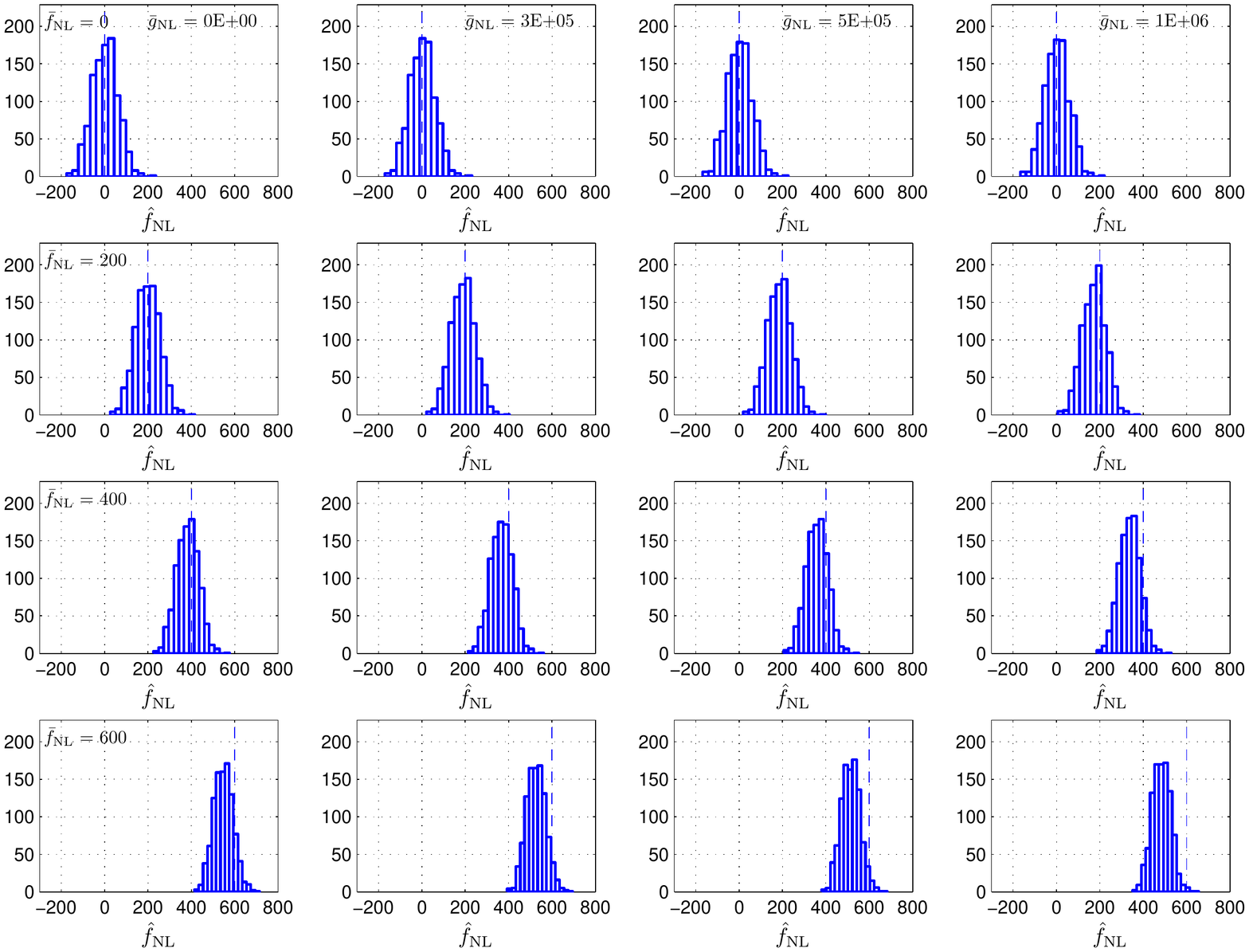}
\caption{\label{fig:fnl_response}These panels represent the accuracy and efficiency on the estimation
of the f$_\nl$ parameter. From left to right, the columns correspond to simulated
$\bar{\rm g}_\nl$ values of: 0, $3\times 10^{5}$, $5\times 10^{5}$ and $10^{6}$.
Similarly, from top to bottom, rows correspond to simulated $\bar{\rm f}_\nl$ values of:
0, 200, 400 and 600. The histograms show the distribution of the obtained values of
$\hat{\rm f}_\nl$ for each case. The vertical dashed lines indicate the simulated $\bar{\rm f}_\nl$ value, and help
to identify the presence of a possible bias.}
\end{center}
\end{figure*}

The results obtained from the 1000 simulations are given in figure~\ref{fig:fnl_response}.
We have explored 16 different non-Gaussian models, accounting for all the possible combinations
obtained with simulated $\bar{\rm g}_\nl$ values of 0, $3\times 10^{5}$, $5\times 10^{5}$ and $10^{6}$, and  $\bar{\rm f}_\nl$
values of 0, 200, 400 and 600. For each panel, we present the histogram of the maximum-likelihood estimation of
the non-linear coupling \emph{quadratic} parameter: $\hat{\rm f}_\nl$. Notice that we refer to a simulated
value of a given non-linear coupling parameter (x$_\nl$), as $\bar{\rm x}_\nl$, whereas that its estimation via
the maximum-likelihood is denoted as $\hat{\rm x}_\nl$. Vertical dashed lines in each panel, indicate the value
of the maximum-likelihood estimation for the parameter.

As it can be noticed from the figure, when the simulated data satisfies the condition of the particular
explored model (i.e., $\bar{\rm g}_\nl \equiv 0$, first column), the f$_\nl$ is accurately and efficiently
estimated, at least for values of $\bar{\rm f}_\nl < 600$. Actually, this is a result that we already obtained in~\cite{vielva09},
which indicates that for $\bar{\rm f}_\nl > 600$, the perturbative model stars to be not valid any longer.

However, when the simulated non-Gaussian maps also contain a significant contribution from a \emph{cubic} term, the bias in the 
determination of the f$_\nl$ parameter stars to be evident already for lower values of the simulated $\bar{\rm f}_\nl$. It is
interesting to notice that, even if the simulated $\bar{\rm g}_\nl$ is large (for instance  $\bar{\rm g}_\nl = 10^6$), we do
not see any significant bias in $\hat{\rm f}_\nl$, for simulated $\bar{\rm f}_\nl$ values lower than 200.

Summarizing, we can infer that for non-Gaussian scenarios with $|\bar{\rm f}_\nl| \lesssim 400$ and
$|\bar{\rm g}_\nl| \lesssim 5\times 10^{5}$, no significant bias on the estimation of a pure \emph{quadratic} term
is found. It is worth mentioning that these range of values are in agreement with predictions from most of the physically motivated
non-Gaussian inflationary models. Notice that, in general, even for the cases in which a bias is observed, the efficiency in
the determination of f$_\nl$ (somehow related to the width of the histograms) is almost unaltered.

\subsection{The recovery of g$_\nl$ in the presence of a \emph{quadratic} term}

\begin{figure*}
\begin{center}
\includegraphics[angle=90,width=15cm]{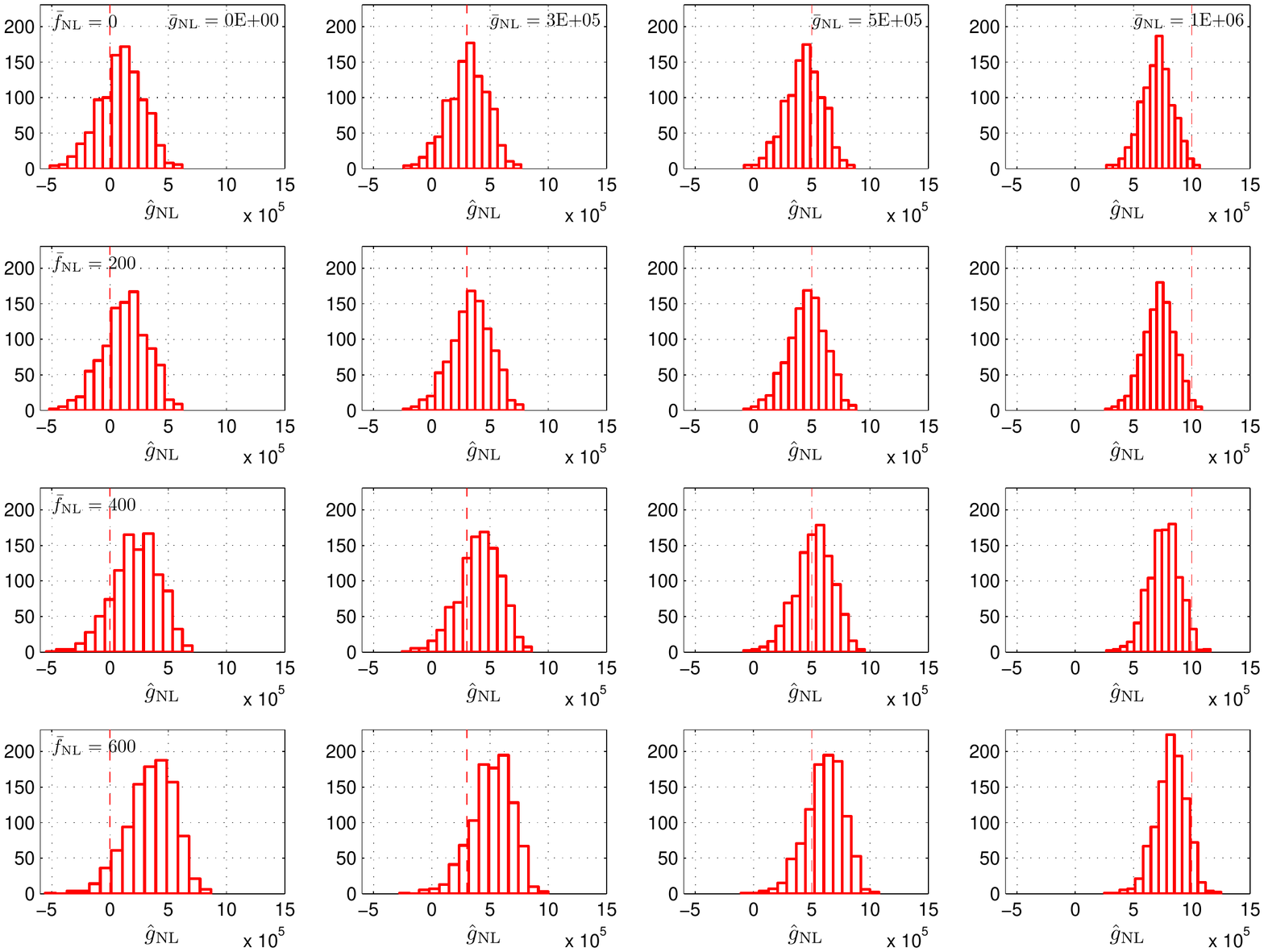}
\caption{\label{fig:gnl_response}These panels represent the accuracy and efficiency on the estimation
of the g$_\nl$ parameter. From left to right, the columns correspond to simulated
$\bar{\rm g}_\nl$ values of: $0$, $3\times 10^{5}$, $5\times 10^{5}$ and $10^{6}$.
Similarly, from top to bottom, rows correspond to simulated $\bar{\rm f}_\nl$ values of:
0, 200, 400 and 600. The histograms show the distribution of the obtained values of
$\hat{\rm g}_\nl$ for each case. The vertical dashed lines indicate the simulated $\bar{\rm g}_\nl$ value, and help
to identify the presence of a possible bias.}
\end{center}
\end{figure*}

As for the previous case, a graphical representation of the results obtained from the 1000 simulations is given in figure~\ref{fig:gnl_response}.
We have explored the same 16 different non-Gaussian models already described above. 
As it can be noticed from the figure, when the simulated data
corresponds to the explored model (i.e., $\bar{\rm f}_\nl \equiv 0$, first row), the g$_\nl$ parameter is reasonably estimated,
at least for simulated $\bar{\rm g}_\nl < 10^6$. 

However, when the simulated non-Gaussian maps also contain a significant contribution from the \emph{quadratic} term, a bias in the 
determination of the g$_\nl$ parameter stars to be notorious for lower values of the simulated $\bar{\rm g}_\nl$ coefficient. In particular, the
plots of the first column (i.e., $\bar{\rm g}_\nl \equiv 0$) show a clear bias on $\hat{\rm g}_\nl$. This
indicates that, when the analyzed case corresponds to a pure \emph{quadratic} scenario, and a pure \emph{cubic} model is assumed, the g$_\nl$
estimator is sensitive to the \emph{quadratic} non-Gaussianity and, somehow, it absorbs the non-Gaussianity in the
form of a fake \emph{cubic} term. In particular, an input value of $\bar{\rm f}_\nl \equiv 400$ is determined as a pure $\hat{\rm g}_\nl \approx 2.5\times 10^5$. Notice that this was not the
situation for the previous case, where the f$_\nl$ estimation was not sensitive to the presence of a pure \emph{cubic} model
(at least for reasonable values of $\bar{\rm g}_\nl$). This is an expected results, since, any skewned distribution would imply the presence of
a certain degree of kurtosis, whereas the opposite is not necessary true.

\subsection{The general case: the joint recovery of f$_\nl$ and g$_\nl$}

\begin{figure*}
\begin{center}
\includegraphics[angle=90,width=15cm]{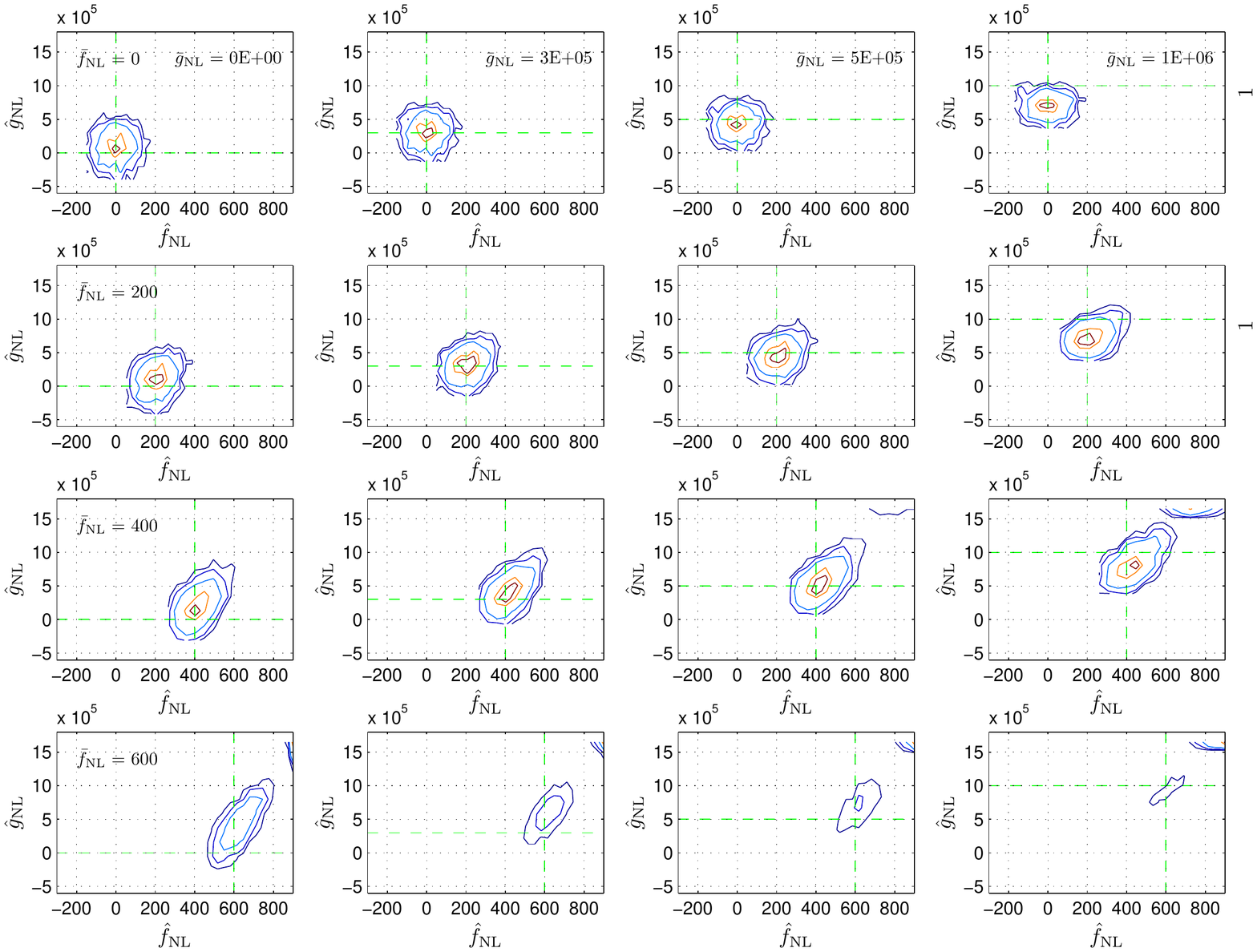}
\caption{\label{fig:fnl_gnl_response}These panels represent the accuracy and efficiency on the joint estimation
of the f$_\nl$ and g$_\nl$ parameters. From left to right, the columns correspond to simulated
$\bar{\rm g}_\nl$ values of: $0$, $3\times 10^{5}$, $5\times 10^{5}$ and $10^{6}$.
Similarly, from top to bottom, rows correspond to simulated $\bar{\rm f}_\nl$ values of:
0, 200, 400 and 600. The contours show the distribution of the obtained values of the
pair $\left( \hat{\rm f}_\nl, \hat{\rm g}_\nl \right)$ for each case. The vertical and horizontal dashed lines indicate the 
simulated $\bar{\rm f}_\nl$ and $\bar{\rm g}_\nl$ values, respectively, and help
to identify the presence of a possible bias.}
\end{center}
\end{figure*}

Finally, we have also explored the case of a joint estimation of the \emph{quadratic} and \emph{cubic} terms. The results obtained from the 1000 simulations
are given in figure~\ref{fig:fnl_gnl_response}.
As for the previous cases, we have explored the same 16 different non-Gaussian models already described above. 
The plots represent the contours of the 2-D histograms obtained for the pair $\left( \hat{\rm f}_\nl, \hat{\rm g}_\nl \right)$ of
the maximum-likelihood estimation. Vertical and horizontal dashed lines indicate the simulated $\bar{\rm f}_\nl$ and
$\bar{\rm g}_\nl$ values, respectively.

As it can be noticed from the figure, only for the regime $|\bar{\rm f}_\nl| \lesssim 400$ and
$|\bar{\rm g}_\nl| \lesssim 5\times 10^{5}$, we obtain an accurate and efficient estimation of the non-linear coupling parameters.
As it was reported above, this regime correspond to the boundaries obtained from the pure f$_\nl$ case.

It is interesting to notice the presence of very large biases for cases outside of the previous range. In particular, estimations tend to
move towards a region of the parameter space of larger values of both, f$_\nl$ and g$_\nl$. Only a secondary peak in the 2-D histogram
corresponds to the simulated pair of values.

This result, combined with the previous ones, indicates that the non-Gaussian model proposed in equation~\ref{eq:physical_model}
is only valid up to values of the \emph{quadratic} and \emph{cubic} terms of around $1\%$ and $0.05\%$, respectively.

\section{Application to WMAP 5-year data}
\label{sec:wmap}

We have studied the compatibility of the WMAP 5-year data with a non-Gaussian model as the
one described in equation~\ref{eq:physical_model}. 
In particular, we
have analyzed the co-added CMB map generated from the global noise-weighted
linear combination of the reduced foreground maps for the Q1, Q2,
V1, V2, W1, W2, W3 and W4 difference assemblies~\citep[see][for
details]{gold09}. The weights are proportional to the inverse average noise variance across
the sky, and are normalized to unity.
This linear combination is made at \nside=512 HEALPix resolution,
being degraded afterwards down to \nside=32.

Under these circumstances, we are in the same condition as for the analysis performed on
the simulations described in Section~\ref{sec:simulations}. Therefore, the theoretical multinormal
covariance of the CMB temperature fluctuations ($\bmath{\xi}$) is the one already computed
with the 500000 simulations (see previous Section).

Two different analysis were performed. The first accounts for an all-sky study (except for the
sky regions covered by the Galactic mask described in the previous Section), where constraints
on the non-linear coupling parameters from different scenarios are presented. We will present
as well results derived from a model selection approach, where we investigate which are the
models that are more favoured by the data. The second analysis explores the
WMAP data compatibility with the local non-Gaussian model in two different hemispheres. In particular,
we have studied independently the two hemispheres related to the dipolar pattern described
in~\cite{hoftuft09}.

\subsection{All-sky analysis}

\begin{figure*}
\begin{center}
\includegraphics[width=5.8cm,keepaspectratio]{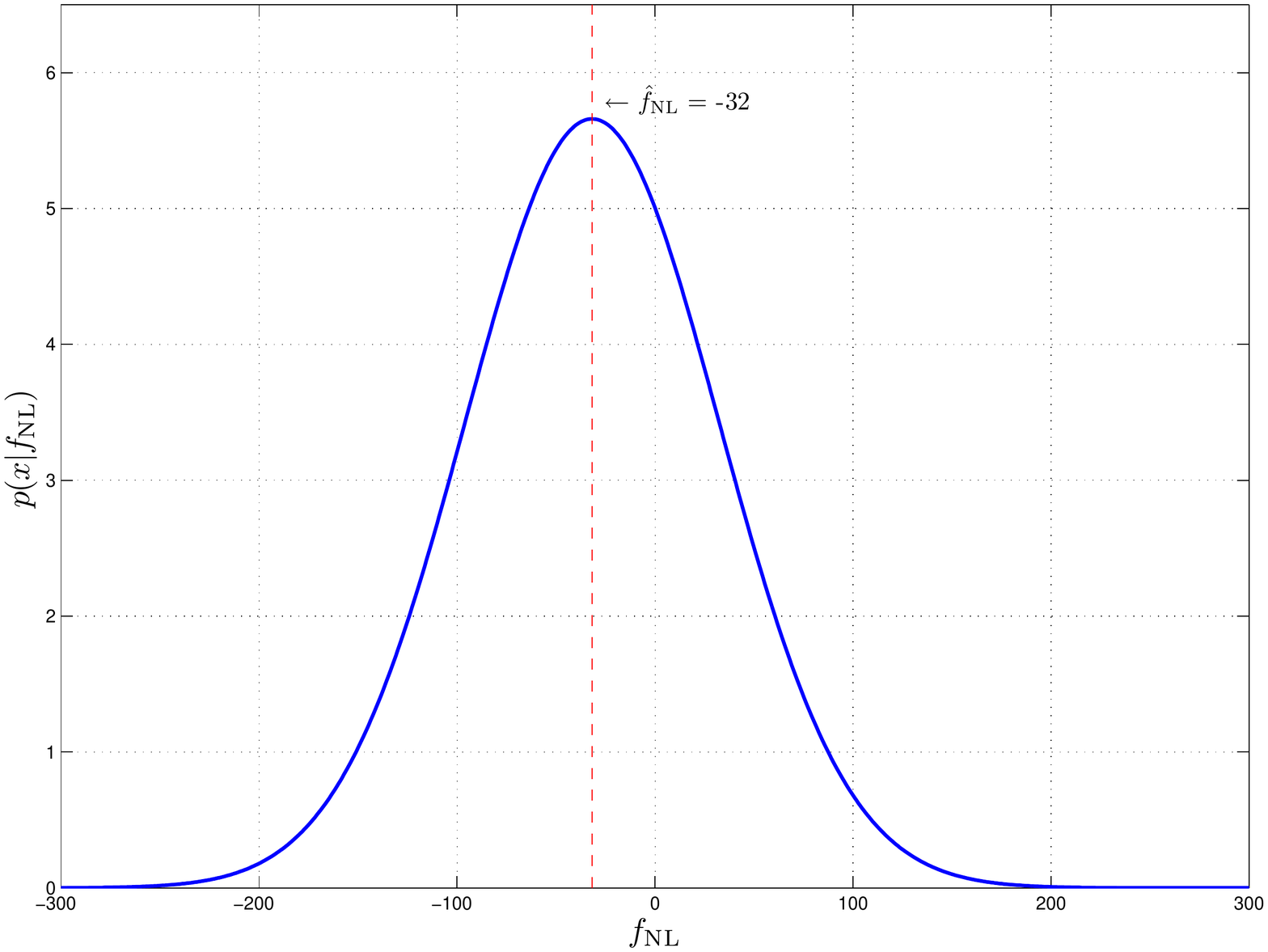}
\includegraphics[width=5.8cm,keepaspectratio]{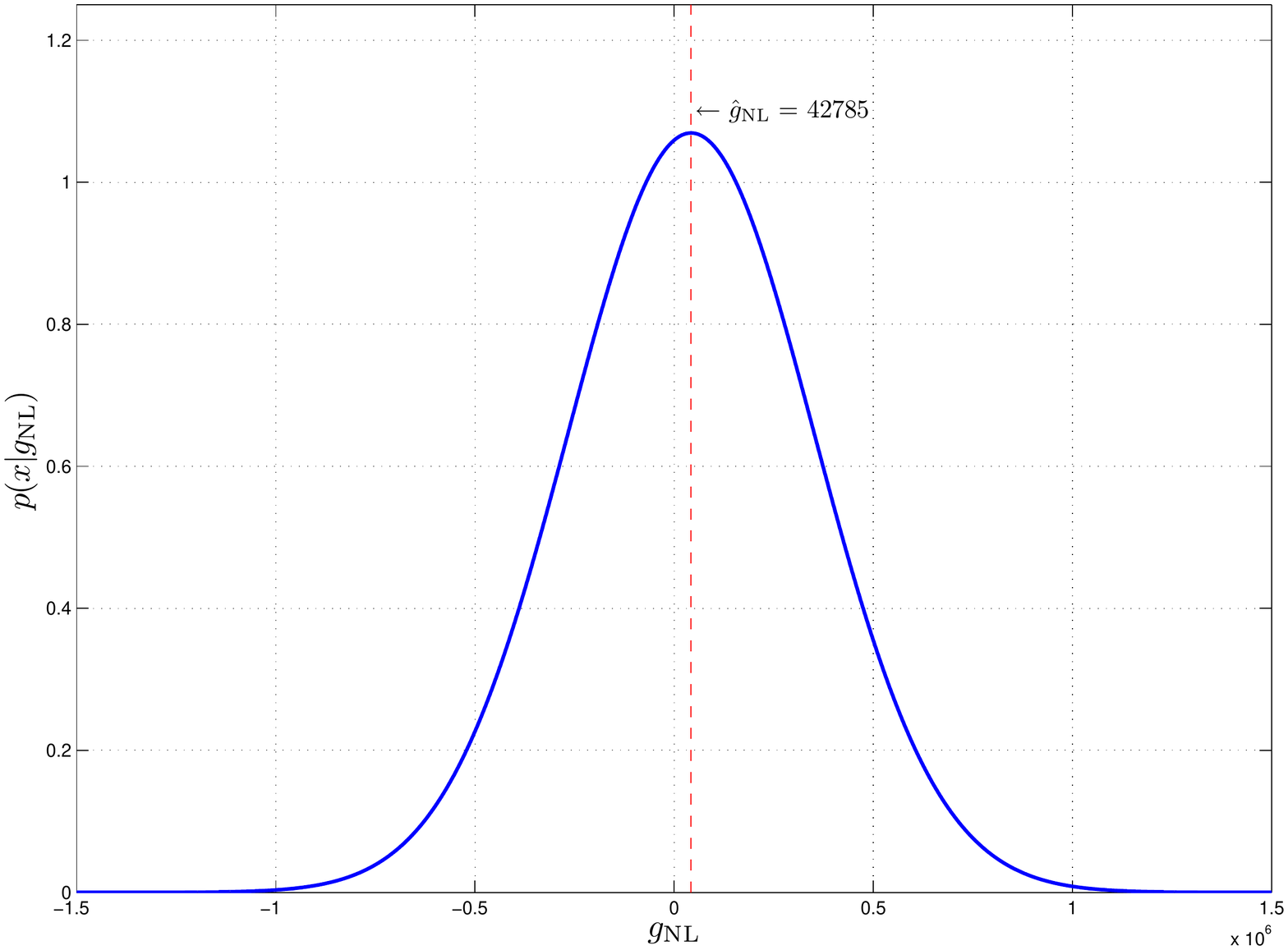}
\includegraphics[width=5.8cm,keepaspectratio]{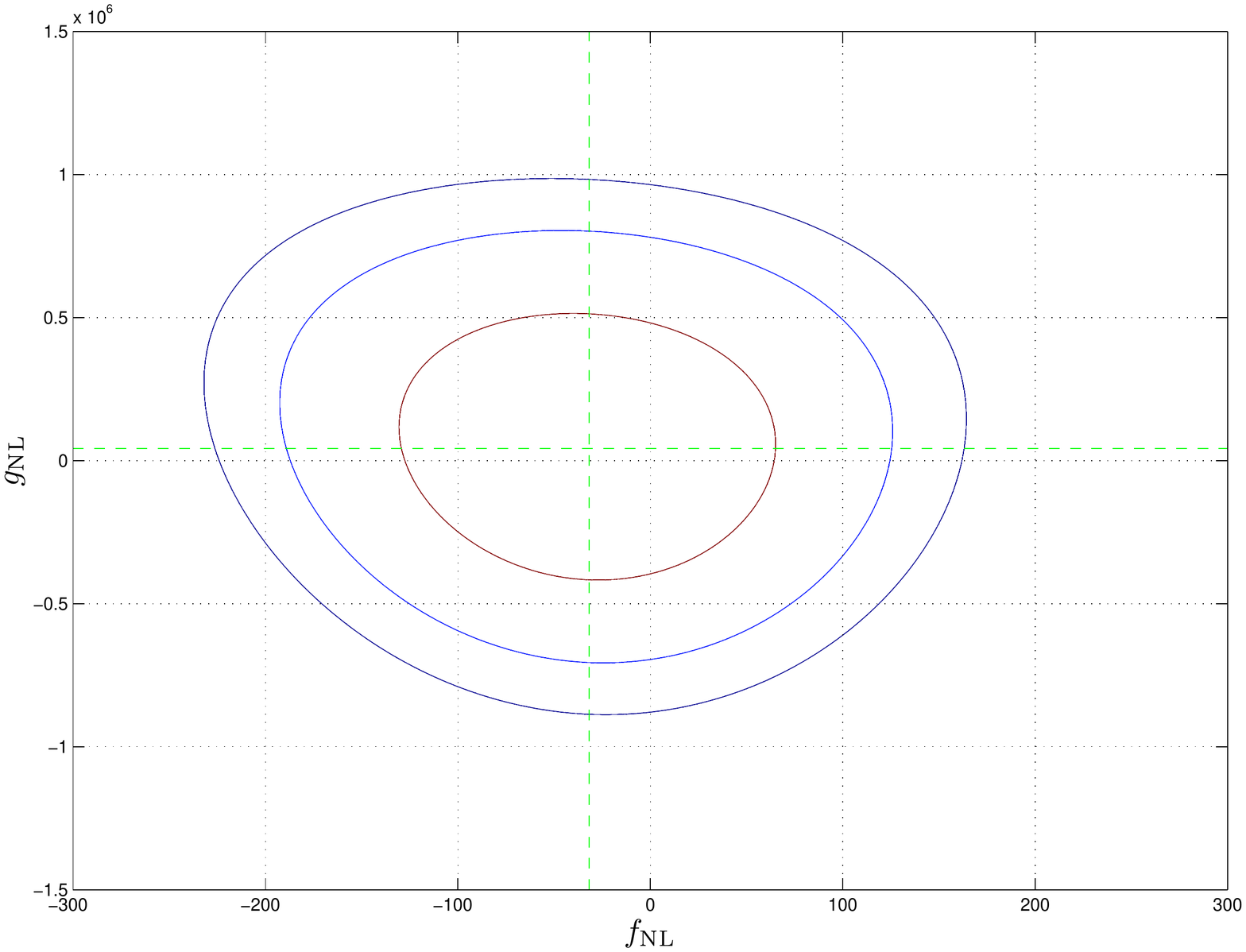}
\caption{\label{fig:data_like}Likelihood distribution
of the non-linear parameters  obtained by analyzing 1000
simulations, according to the local non-Gaussian model given in equation~\ref{eq:physical_model}.
Left panel correspond to a pure f$_\nl$ analysis: $p\left(\bmath{x} | {\rm f}_\nl \right)$. Middle plot shows
the result for a pure g$_\nl$ model:  $p\left(\bmath{x} | {\rm g}_\nl \right)$. Right panel provides the 68\%, 95\% and 99\%
contour levels of the likelihood obtained from a joint f$_\nl$, g$_\nl$ analysis of the WMAP 5-year data:
$p\left(\bmath{x} | {\rm f}_\nl, {\rm g}_\nl  \right)$.}
\end{center}
\end{figure*}

We have computed the full N-pdf in 
equation~\ref{eq:pdfx}, for three different scenarios: a non-Gaussian model with a pure
\emph{quadratic} term (i.e., g$_\nl \equiv 0$), another case with a pure \emph{cubic} term
(i.e., f$_\nl \equiv 0$), and a general non-Gaussian model (i.e., f$_\nl \neq 0$ and g$_\nl \neq 0$),

Results are given in figure~\ref{fig:data_like}. Left panel shows the likelihood obtained for the first
case: $p\left(\bmath{x} | {\rm f}_\nl \right)$. Actually, this result is the one
that we already obtained in our previous work~\citep{vielva09}. The maximum-likelihood estimation for the
\emph{quadratic} factor is $\hat{{\rm f}}_\nl = -32$\footnote{Notice that in~\cite{vielva09} we used a different
definition between the primordial gravitational potential $\Phi$ and the CMB temperature. This difference implies that,
in our previous work, the f$_\nl$ parameter had an opposite sign with respect to the definition used in
this paper.}. The constraint on the non-Gaussian parameter
is: $-154 < {\rm f}_\nl < 94$ at 95\%.

The middle panel in figure~\ref{fig:data_like} presents the likelihood obtained from a model with a pure \emph{cubic} term:
$p\left(\bmath{x} | {\rm g}_\nl \right)$. The maximum-likelihood estimation for the
\emph{quadratic} factor is $\hat{{\rm g}}_\nl = 42785$. The constraint on the parameter is: 
$-5.6\times 10^5 < {\rm g}_\nl < 6.4\times 10^5$ at 95\%. This result is compatible with a previous finding
obtained from the analysis of LSS data~\citep{desjacques09}. The result reported in this work is, as far as we know, the first direct
constraint of g$_\nl$ from CMB data alone.

The right panel in figure~\ref{fig:data_like} shows the contour levels at the 68\%, 95\% and 99\% CL, for the
likelihood obtained from an analysis of a general \emph{quadratic} and \emph{cubic} model: $p\left(\bmath{x} | {\rm f}_\nl, {\rm g}_\nl  \right)$. 
Notice that the maximum likelihood estimation
for the f$_\nl$ and g$_\nl$ parameters are similar to those obtained from the previous cases (where the pure models
were investigated). Even more, the marginalized distributions for the two parameters are extremely similar to the 
likelihood distributions discussed previously, and, therefore, the constraints on the non-linear coupling
parameters are virtually the same.

Finally, we want to comment a few words about two issues: the incorporation of possible \emph{a priori} information related to the parameters
defining the non-Gaussian model, and the application of model selection criteria (or hypothesis tests) to discriminate 
among the
Gaussian model and different non-Gaussian models.

As we largely discussed in our
previous work~\citep{vielva09}, one of the major advantages of computing the full N-pdf on the non-Gaussian model is that, 
in addition to provide a maximum-likelihood estimation for the non-linear coupling parameters, we have a full description
of the statistical properties of the problem.
More in particular, if we could have any physical (or empirical) motivated prior for the f$_\nl$ and g$_\nl$ parameters, it
could be used together with the likelihood function to perform a full Bayesian parameter estimation. 
This aspect has not
been considered in this work, precisely because such a well motivated prior is lacking. 
Actually, a possible and trivial \emph{a priori} information that
could be used in this specific case, would be to limit the range of values that can be taken by
f$_\nl$ and g$_\nl$, such as the non-Gaussian model is, indeed, a 
local perturbation of a Gaussian field (i.e., the typical values that we discussed in Section~\ref{sec:simulations}). 
However, these priors do not seem to be quite useful since, first, we do not have any evidence
to chose any different form for the prior that an uniform value over the parameters range; and, second, the limits of these ranges are
somehow arbitrary. These kind of priors do not provide any further knowledge on the Bayesian parameter
determination: as it is well known, such estimation would be totally driven by the likelihood itself, since it is fully defined within
any reasonable \emph{a priori} ranges.

The possibility of performing a model selection approach is an extra advantage of dealing with the full N-pdf. 
Of course, under the presence of an hypotetical well motivated prior on the non-linear coupling
parameters, model selection could be done in terms of the Bayesian evidence or the ratio of posterior 
probabilities~\citep[see][for a specific discussion on this application]{vielva09}.
However, the lack of such a prior (as we discussed above), makes the application of a full Bayesian model selection
framework significantly less powerful than in other situations:
as it is very well known, the use of uniform priors for all the
parameters would provide very little information,
since the results would be very much dependent on the size of the parameters range.
Despite this, we can still make a worthy use of the likelihood to perform model selection.
In particular, some asymptotic model selection criteria, like the \emph{Akaike Information Criteria}~\citep[AIC,][]{akaike73} and the
\emph{Bayesian Information Criteria}~\citep[BIC,][]{schwarz78}, can be applied. Both methods provide a ranging index for competitive hypotheses, where the
most likely one is indicated by the lowest value of the index. The AIC and BIC indices depend on the maximum value of
the log-likelihood ($\max  \left[ {\mathcal L} \left(\bmath{x} | \Theta \right) \right] \equiv \hat{{\mathcal L}}$):
\begin{eqnarray}
{\rm AIC} \left( H_i \right) & = & 2\left(p - \hat{{\mathcal L}}\right), \nonumber \\
{\rm BIC} \left( H_i \right) & = & 2\left(\frac{p}{2}\log N - \hat{{\mathcal L}}\right), \nonumber
\end{eqnarray}
where $p$ is the number of parameters that determine the hypothesis or model $H_i$. We have applied these two asymptotic
model selection criteria to the WMAP 5-year data. Defining the Gaussian model as $H_0$, the pure \emph{quadratic} model as
$H_1$, the pure \emph{cubic} model as $H_2$, and the general non-Gaussian model as $H_3$, and considering the maximum
value for the log-likelihoods obtained for all these cases, we obtain: 
${\rm AIC} \left( H_0 \right) <  {\rm AIC} \left( H_1 \right) < {\rm AIC} \left( H_2 \right) <  {\rm AIC} \left( H_3 \right)$,
and
${\rm BIC} \left( H_0 \right) <  {\rm BIC} \left( H_1 \right) < {\rm BIC} \left( H_2 \right) <  {\rm BIC} \left( H_3 \right)$.
That is, the most likely model is the Gaussian one (what is in agreement with the results obtained from the 
parameter determination, since f$_\nl \equiv 0$ and g$_\nl \equiv 0$
can not be rejected at any meaningful confidence level). Among the non-Gaussian models, a pure f$_\nl$ model is the most likely
scenario, being a joint f$_\nl$, g$_\nl$ model the most disfavoured by the WMAP 5-year data.

\subsection{Hemispherical analysis}

\begin{figure*}
\begin{center}
\includegraphics[width=8cm,keepaspectratio]{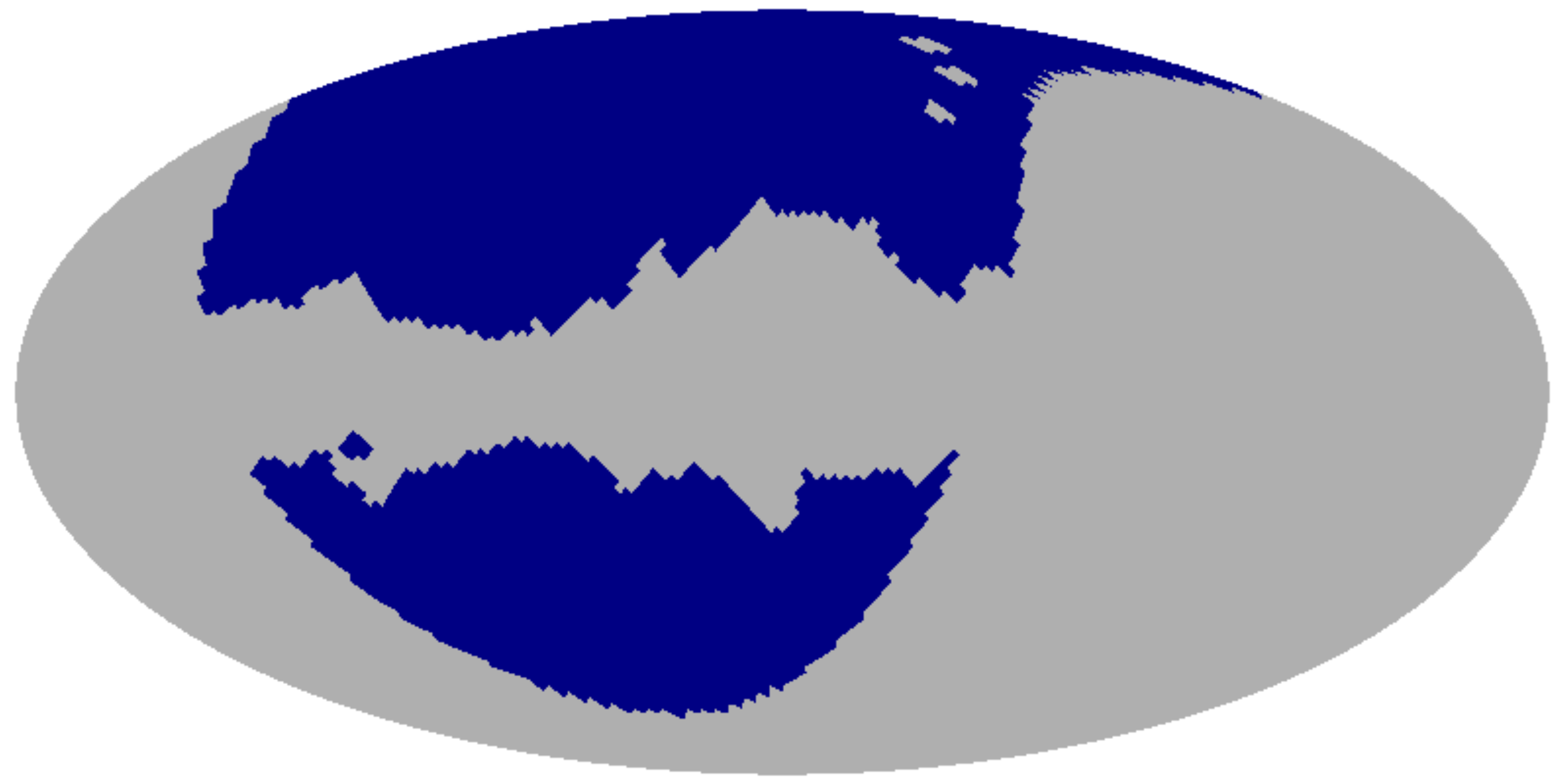}
\includegraphics[width=8cm,keepaspectratio]{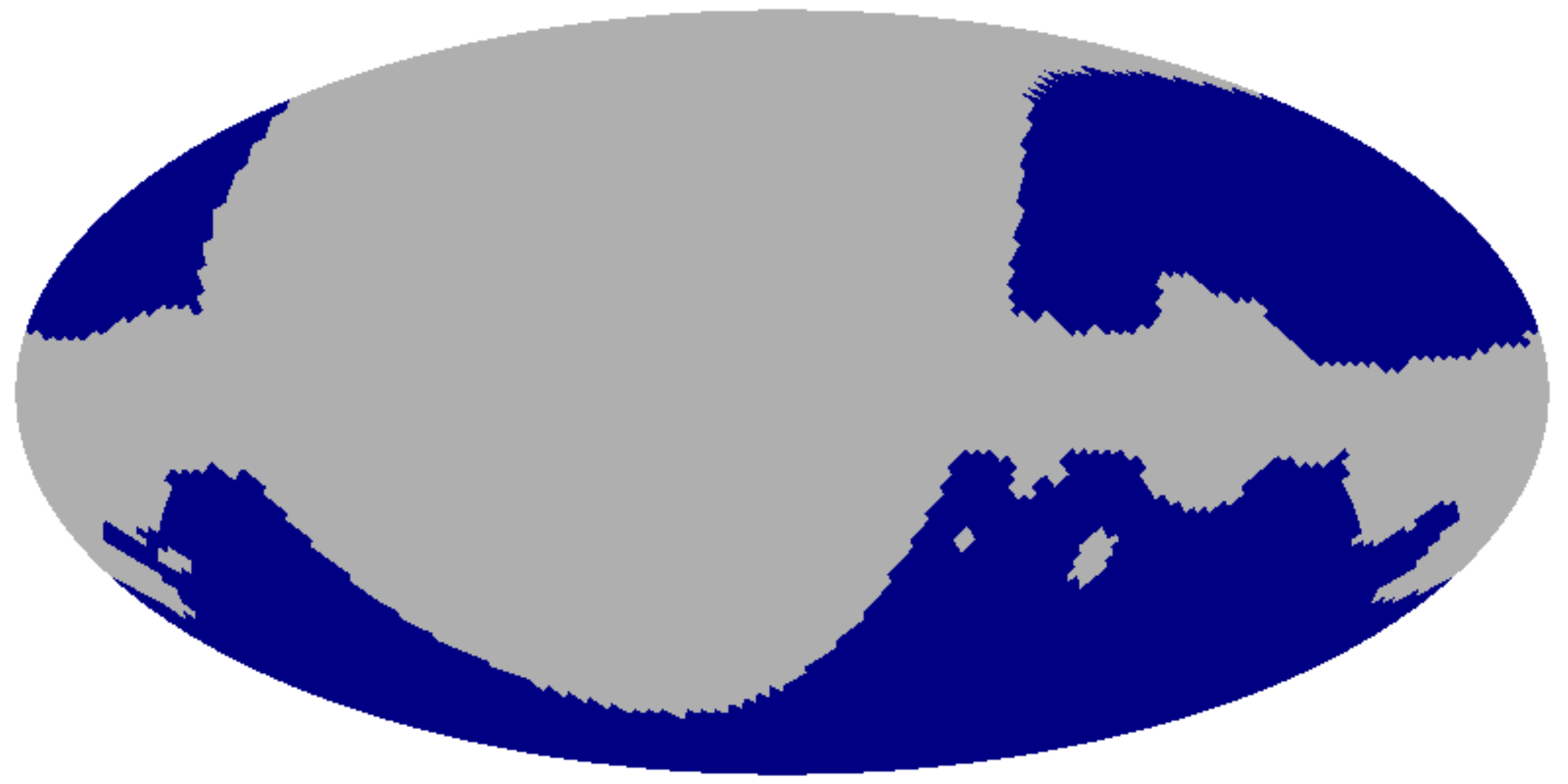}
\caption{\label{fig:dipole_sky}These plots show the areas of the sky that are
independently analyzed. The panel on the left accounts for the sky that, being
allowed by the Galactic mask (see Section~\ref{sec:simulations}), corresponds to the
\emph{Northern hemisphere} of the sky division considered in this work.
Equivalently, the right panel presents the region of the sky that is
analyzed when the \emph{Southern hemisphere} is addressed.}
\end{center}
\end{figure*}

Among the large number of the WMAP \emph{anomalies} that have been reported in the
literature, an anisotropy manifested in the form of a hemispherical asymmetry is one
of the topics that has been more largely studied~\citep[e.g.,][]{eriksen04a,hansen04b}. Most of the works related to
this issue, have reported that such asymmetry is more marked for a north-south
hemispherical division relatively close to the Northern and Southern Ecliptic hemispheres.

In a recent work,~\cite{hoftuft09} reported that large scale WMAP data was compatible with
such kind of anisotropy, in the form of a dipolar modulation defined by
a preferred direction pointing toward the Galactic coordinates $\left(l, b \right) =
\left(224^\circ, -22^\circ \right)$.

Motivated by these results, we have repeated the analysis described in the previous
subsection, but addressing independently the two hemispheres associated to
the dipolar pattern found by~\cite{hoftuft09}. Hereinafter, we will refer to
the \emph{Northern hemisphere} of this dipolar pattern, as the half the of celestial sphere
whose pole is closer to the Northern Ecliptic Pole, and, equivalently, we will
indicate as the \emph{Southern hemisphere} the complementary half of the sky.
The corresponding areas of the sky that are analyzed are shown in figure~\ref{fig:dipole_sky}.
The left and right panels show the allowed sky regions, when the \emph{Northern} and
\emph{Southern hemispheres} of the dipolar pattern are independently addressed, respectively.
Notice that the regions not allowed by the Galactic mask are also excluded from the analysis.
The portion of the sky that is analyzed is around 34\% for the \emph{Northern hemisphere}, and
around 35\% for the \emph{Southern} half.

\begin{figure*}
\begin{center}
\includegraphics[width=8cm,keepaspectratio]{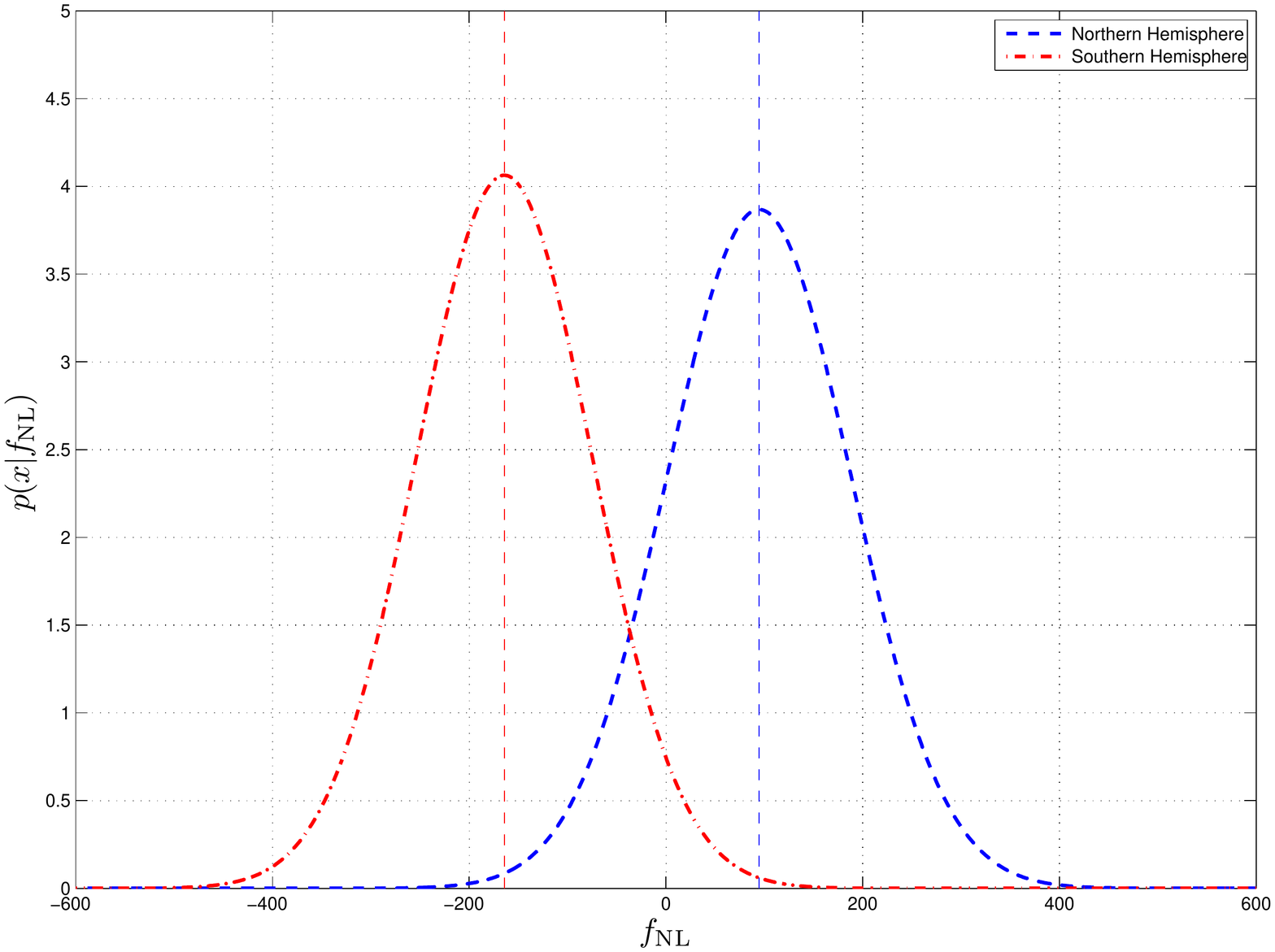}
\includegraphics[width=8cm,keepaspectratio]{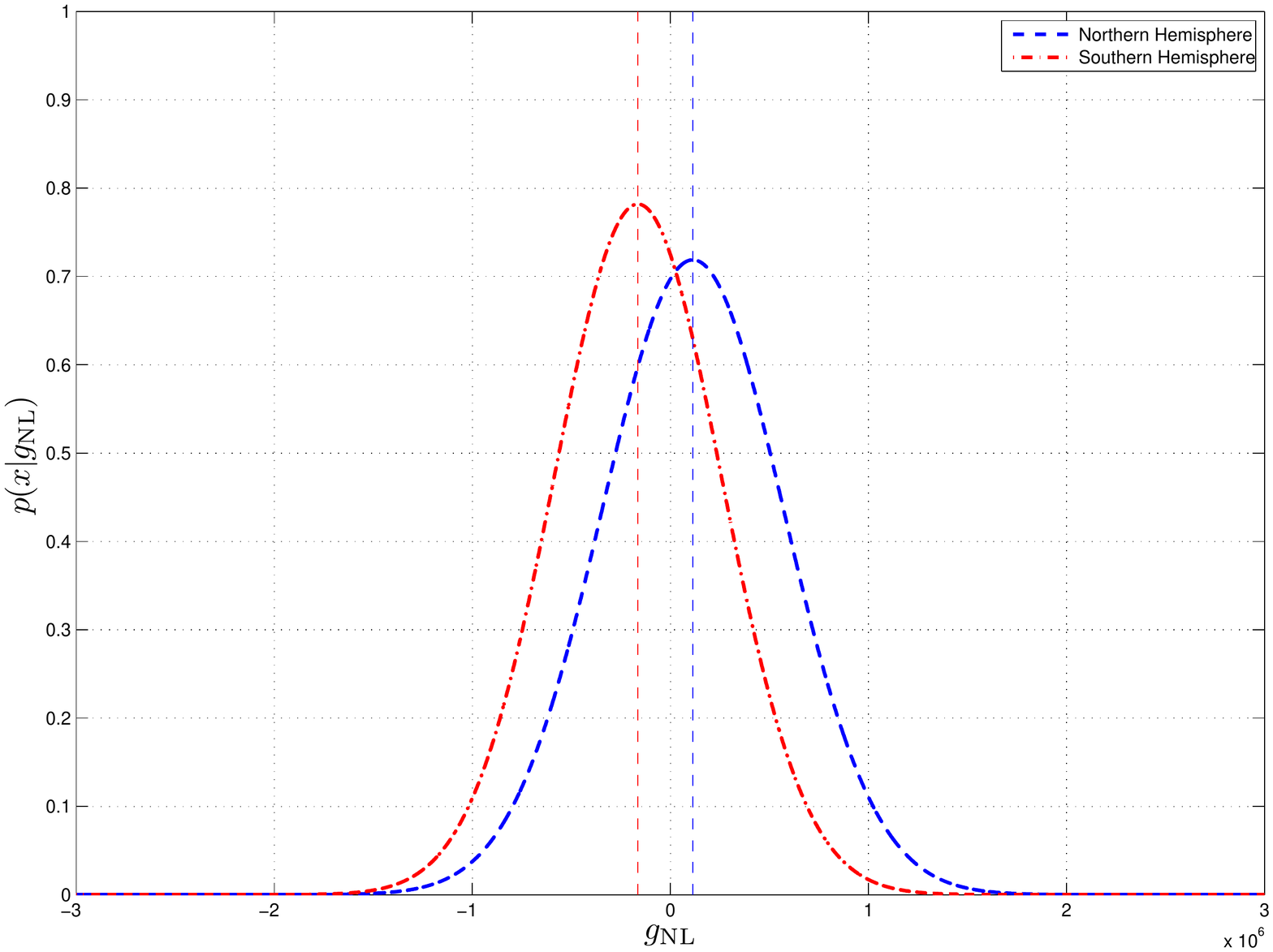}
\caption{\label{fig:dipole_posterior}The left panel present the likelihood on f$_\nl$
obtained from a pure \emph{quadratic} analysis ---$p\left(\bmath{x} | {\rm f}_\nl \right)$---, whereas
the right plot provides the likelihood on g$_\nl$ from a pure \emph{cubic} study ---$p\left(\bmath{x} | {\rm g}_\nl \right)$.
Dashed lines correspond to the \emph{Northern hemisphere}, whereas dot-dashed lines are
for the \emph{Southern} half. Vertical lines indicate the maximum likelihood
estimation of the non-linear coupling parameters: $\hat{\rm f}_\nl$ and $\hat{\rm g}_\nl$.}
\end{center}
\end{figure*}

As we discussed in the previous subsection, the constraints of the f$_\nl$ and g$_\nl$
parameters obtained from the analysis of pure \emph{quadratic} and \emph{cubic} non-Gaussian
models, do not differ significantly from those obtained from a general analysis of
a joint scenario. This is expected for a regime of relatively low values of the non-linear coupling parameters.
For that reason, in the present study we will only consider the following two
cases: a pure \emph{quadratic} (i.e. g$_\nl \equiv 0$) and a pure \emph{cubic} (i.e. f$_\nl \equiv 0$) models.
Results are given in figure~\ref{fig:dipole_posterior}. We present the likelihood
probabilities for the first case ---$p\left(\bmath{x} | {\rm f}_\nl \right)$--- in the left
plot, and the one corresponding to the second case ---$p\left(\bmath{x} | {\rm g}_\nl \right)$---
in the right panel. Each plot shows the results for the \emph{Northern} (dashed lines), and the \emph{Southern}
(dot-dashed lines) hemispheres. The maximum likelihood estimation for the non-linear coupling parameters are
given as vertical lines.

The right panel shows that both hemisphere have a similar likelihood ($p\left(\bmath{x} | {\rm g}_\nl \right)$)
for the case of a pure \emph{cubic} model. However, interestingly, it is not the case when addressing a
g$_\nl \equiv 0$ scenario. In this case, we notice two important results. First, whereas the f$_\nl$ estimation from
the analysis of the \emph{Northern hemisphere} provides a constraint compatible
with the Gaussian scenario, it is not the case for the \emph{Southern hemisphere}. In fact, we find that
f$_\nl < 0$ at 96\% CL. In particular we find: $\hat{\rm f}_\nl = -164 \pm 62$. Second, the distance between both distributions 
is too large. Let us make use of the Kullback--Leibler divergence~\citep[KLD,][]{kullback51} as a measurement
of the distance between the two likelihoods $p_n\left(\bmath{x} | {\rm f}_\nl \right)$ and $p_s\left(\bmath{x} | {\rm f}_\nl \right)$:
\begin{equation}
D_{n,s} = \int {\rm d}{\rm f}_\nl p_n\left(\bmath{x} | {\rm f}_\nl \right) \log{\frac{p_n\left(\bmath{x} | {\rm f}_\nl \right)}{p_s\left(\bmath{x} | {\rm f}_\nl \right)}},
\end{equation}
where $p_n\left(\bmath{x} | {\rm f}_\nl \right)$ and $p_s\left(\bmath{x} | {\rm f}_\nl \right)$ are the likelihoods for the 
\emph{Northern} and the \emph{Southern hemispheres}, respectively.
Actually, we use the symmetrized statistitic $D$, defined as:
\begin{equation}
D = \frac{1}{2}\left( D_{n,s} + D_{s,n} \right).
\end{equation}
We have found that
the distance $D$ for the likelihood distributions of the f$_\nl$ parameter estimated
in the \emph{Northern} and the \emph{Southern hemispheres} defined by the dipolar pattern described by~\cite{hoftuft09}
is much larger than it would be expected from Gaussian and isotropic random CMB simulations. In particular,
such a distance has a p-value $\approx 0.04$. This result is a further evidence on the largely discussed WMAP
North-South asymmetry, and it is as well an indication that such asymmetry is manifested in terms of the non-Gaussianity
of the CMB temperature fluctuations, in agreement with previous 
results~\citep[e.g.,][]{park04,hansen04b,eriksen04b,vielva04,cruz05,eriksen05,land05,monteserin08,rath09,rossmanith09}.

\begin{figure*}
\begin{center}
\includegraphics[width=8cm,keepaspectratio]{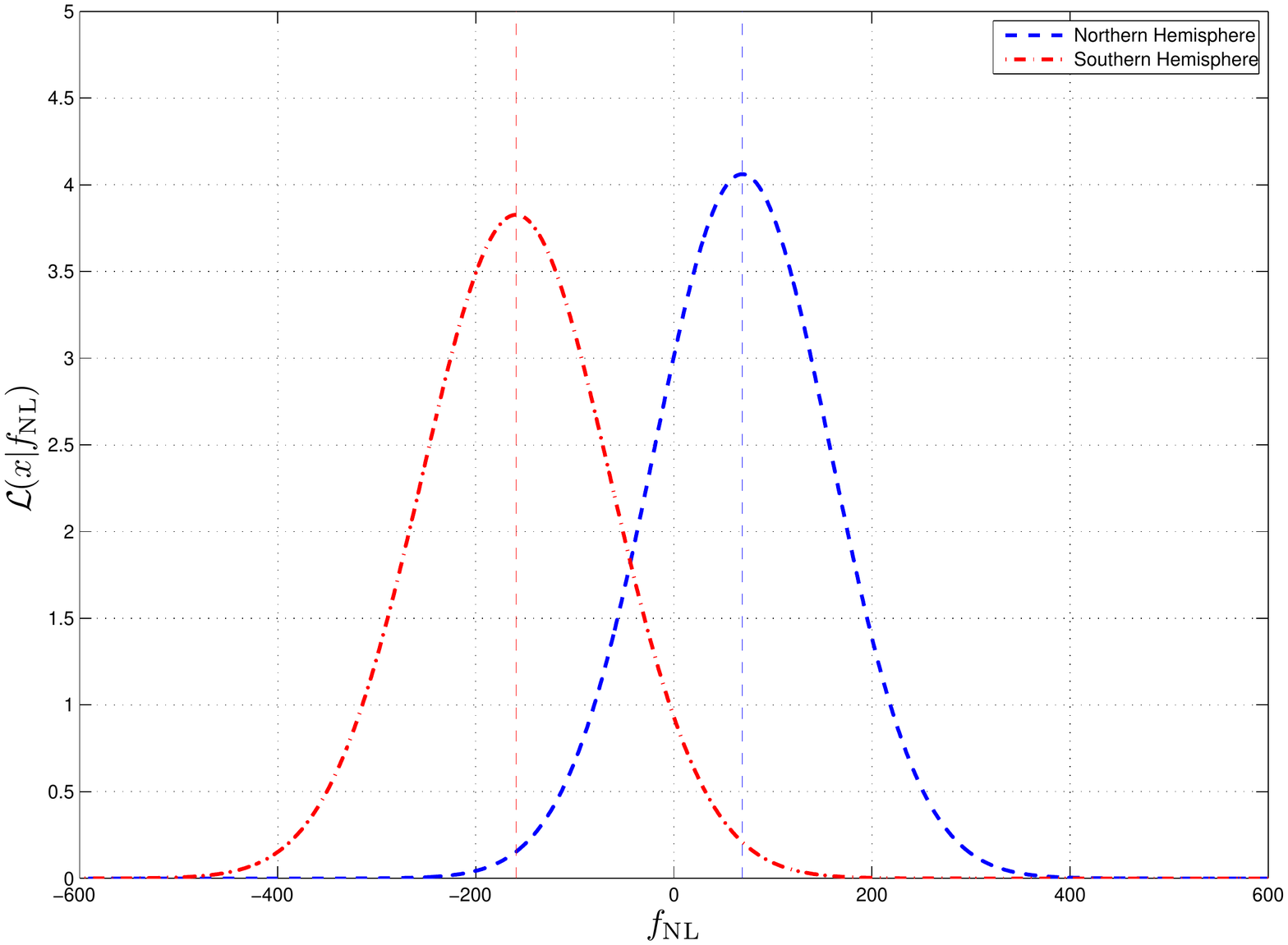}
\includegraphics[width=8cm,keepaspectratio]{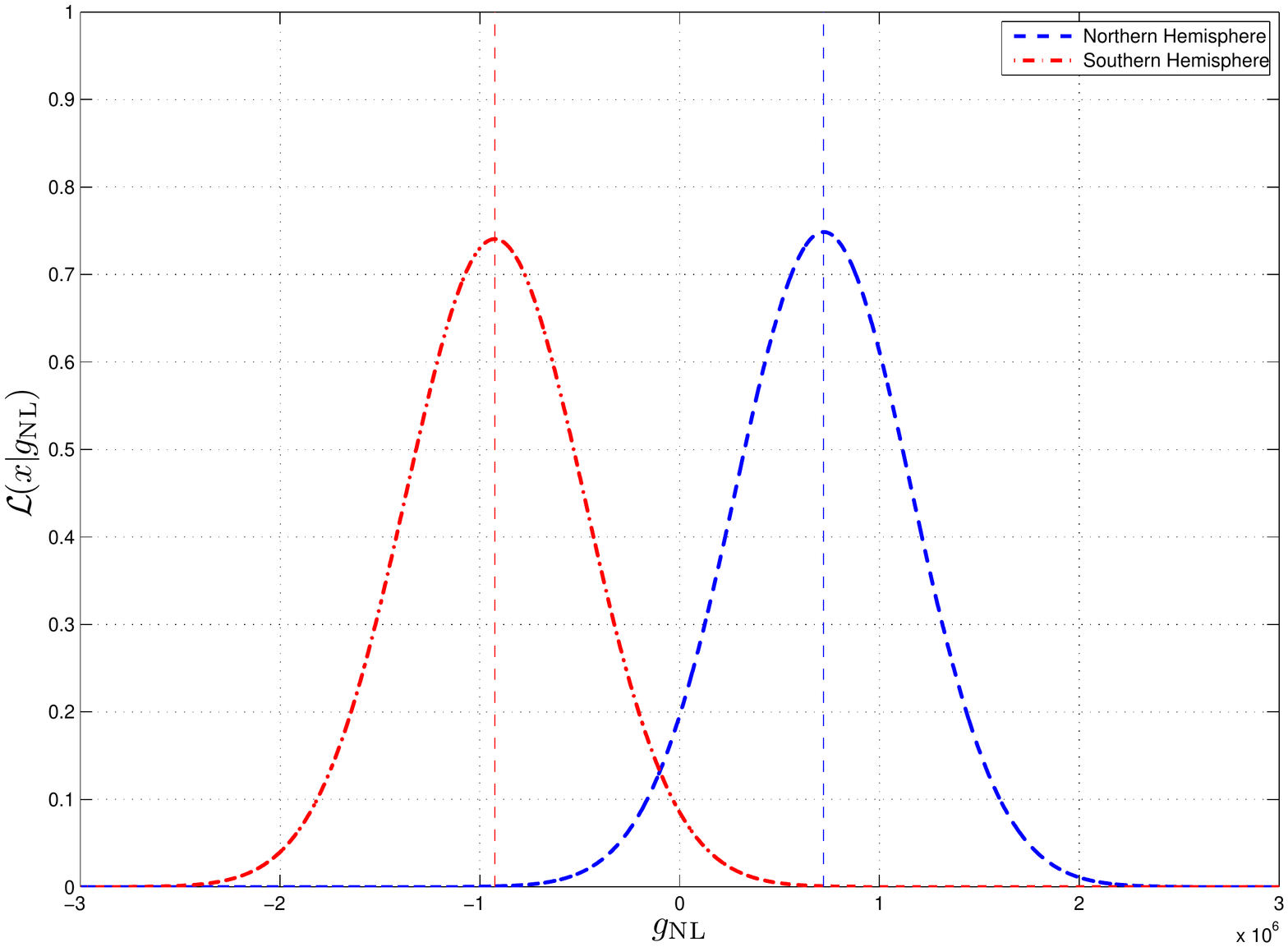}
\caption{\label{fig:modulated_sims} As in figure~\ref{fig:dipole_posterior}, but for the case
in which the WMAP data has been analyzed using correlation matrices that account for the dipolar modulation.}
\end{center}
\end{figure*}

At this point, it is worth recalling that the analysis described above has been performed assuming isotropy, i.e., we have used the same type of correlations to described the second-order statistics in both the \emph{Northen} and in the \emph{Southern hemispheres}.
However, the result obtained by~\cite{hoftuft09} indicates that these two hemisphere might be described by two different correlations (i.e., the sky would not be isotropic any longer).
The dipolar modulation proponed by~\cite{hoftuft09} was small (its amplitude was lower than 0.7\%), but significant (a $3.3\sigma$ detection was claimed).
Assuming this point, we have repeated our previous analysis, but using different statistical properties for the correlation matrices in the two halves.
The way we have estimated these new correlation matrices is as follows: we have generated 500,000 simulations (in the same way it has been already described at the beginning of Section~\ref{sec:wmap}, and, once the co-added maps are degraded to \nside=32, each one of the simulations have been modified by applying the dipolar modulation estimated by~\cite{hoftuft09} from the WMAP data.
It is from these modulated simulations that we have estimated the new correlation matrices needed to estimate the likelihood probabilities. The result of this test is presented in figure~\ref{fig:modulated_sims}. The conclusions related to the f$_\nl$ estimation are essentially the same: on the one hand the analysis of the \emph{Northern hemisphere} provides a constraint compatible with the Gaussian scenario, whereas the \emph{Southern hemisphere} is not; on the other hand, the distance between both distributions is again too large (it also has a p-value --as compared to, in this case, anisotropic simulations-- increases up to $\approx 0.09$ (estimated in terms of the KLD). 
However, despite this slight change in the f$_\nl$ hemispherical estimation, dramatic differences can be observed for the pure \emph{cubic} scenario.
Interestingly, accounting for the dipolar modulation correction reveals an extra departure form anisotropy related to the g$_\nl$ constraints. The dipolar modulation makes the maximum likelihood estimation of g$_\nl$ highly incompatible between both hemispheres. In particular,  the distance between both distributions is extremely rare as compared with the expected behaviour from 
Gaussian and anisotropic CMB simulations (generated, as explained above, by applying the dipolar modulation reported by~\cite{hoftuft09}): it has a p-value of $\approx 0.002$, in the sense of the KLD.

\begin{figure*}
\begin{center}
\includegraphics[width=8cm,keepaspectratio]{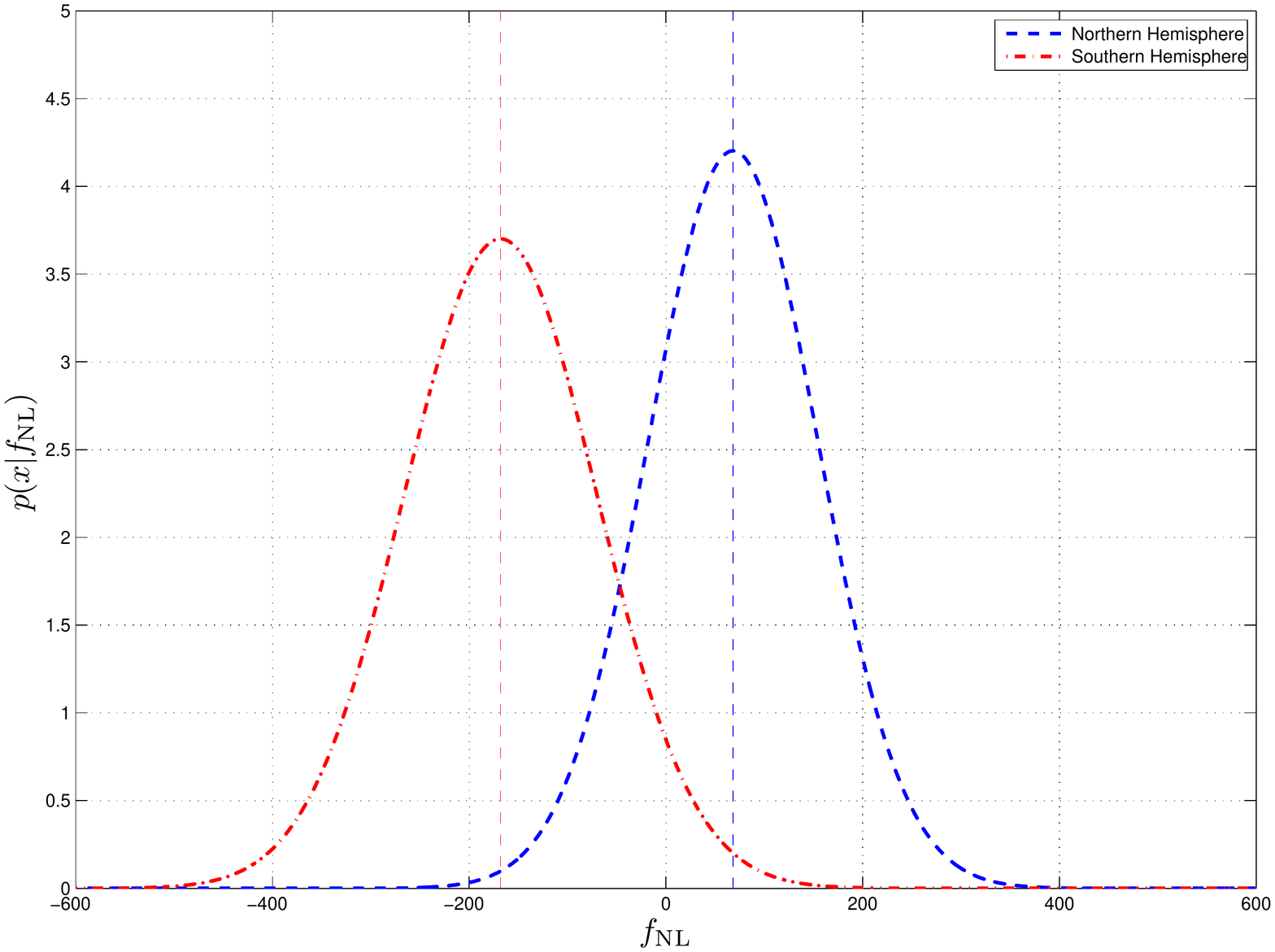}
\includegraphics[width=8cm,keepaspectratio]{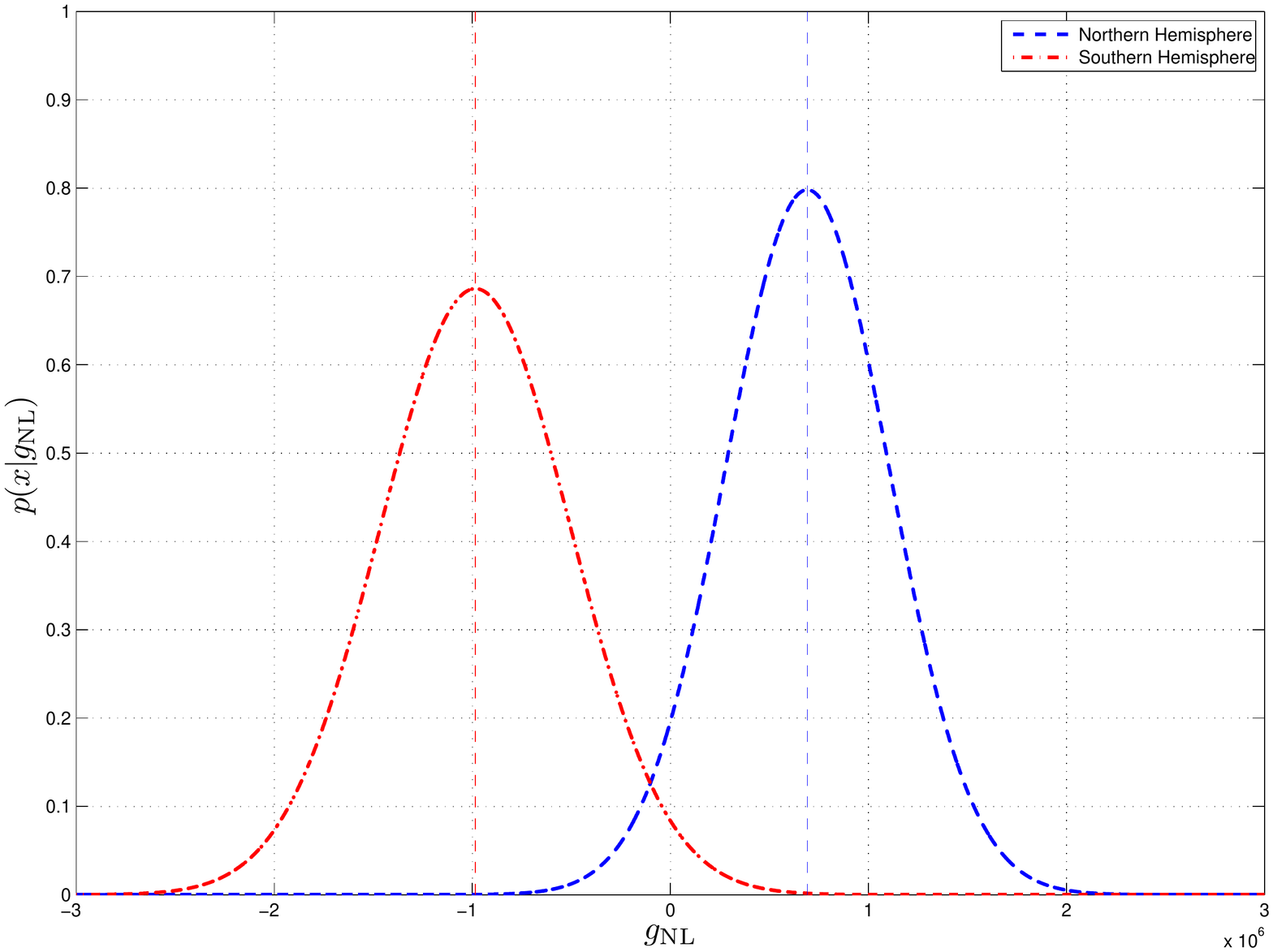}
\caption{\label{fig:dipole_posterior_corrected} As in figure~\ref{fig:dipole_posterior}, but for the case
in which the WMAP data has been corrected by the dipolar modulation.}
\end{center}
\end{figure*}

We have also studied whether a WMAP data corrected by the dipolar modulation found by~\cite{hoftuft09} could
present a behaviour compatible with the Gaussian and isotropic hypotheses. Results for the corrected WMAP data are given in figure~\ref{fig:dipole_posterior_corrected}.
Notice that we do not see any
significant differences from the previous situation (i.e., the case in which uncorrected WMAP data was analyzed assuming anisotropy): the 
dipolar modulation does not affect
to the f$_\nl$ and g$_\nl$ constraints.

Summarizing, the results obtained in this subsection seem to confirm that there is some kind of anomaly related to an hemispherical asymmetry as the one defined by the dipolar pattern reported by~\cite{hoftuft09}, in the sense of the f$_\nl$ parameter. Even more, when WMAP data is analyzed using correlations compatible with the dipolar modulation suggested by~\cite{hoftuft09}, not only asymmetries related to the f$_\nl$ parameter are clear, but also associated to the \emph{cubic term} (i.e., the g$_\nl$ parameter). Intriguingly, the correction of the WMAP data in terms of this dipolar modulation is not enough to obtain a CMB signal compatible with a Gaussian and isotropic random field.

At this point, it is worth mentioning that the dipolar modulation of~\cite{hoftuft09}
was obtained by considering second-order moments of the CMB data and, therefore, this correction only
addresses the problems related to an asymmetry in terms of this order.
Hence, it is not totally surprising that this dipolar modulation correction is not sufficiently satisfactory to
solve the anomaly reported in this work, since such anomaly is related to higher-order moments.
It is also need to point out that we have tested that the dipolar modulation correction of the WMAP data does not affect the
results obtained from an all-sky analysis of the CMB data.

Finally, let us recall that, in a recent work,~\cite{rudjord09} searched for specific asymmetries related to the local estimation of the f$_\nl$ parameter, by using needlets. Contrarily to our findings, in this work no significant asymmetry was found when analyzing WMAP data. There are some differences between the analyses that could explain the discrepancy, although they have to be taken as mere suggestions. 
First, the kind of non-Gaussianity that is probed by each work is different: whereas the~\cite{rudjord09} paper explore a f$_\nl$ model that is local in the gravitational potential (from which the non-Gaussian temperature fluctuations are obtained taking into account all the gravitational effects), here we adopt a local model in the Sachs-Wolfe regime. Second, they work at the best WMAP resolution (around 10-20 arcmins), whereas we focus on scales of around $2^\circ$. Third, we explore an specific division of the sky (the one reported by~\cite{hoftuft09}), whereas they consider several divisions that, not necessarily, have to match the one used by us (they explore hemispherical divisions within an interval of around $30^\circ$).

\section{Conclusions}
\label{sec:final}

We have presented an extension of our previous work~\citep{vielva09}, by
defining a parametric non-Gaussian model for the CMB temperature
fluctuations.
The non-Gaussian model is a local perturbation of the standard CMB Gaussian field,
which (under certain circumstances) is related to 
an approximative form of the weak non-linear coupling inflationary model
at scales larger than the
horizon scale at the recombination time 
\citep[i.e., above the degree
scale, see for instance][]{komatsu01,liguori03}.
For this model, we are able to build the posterior probability of the data given the non-linear
parameters f$_\nl$ and g$_\nl$. From these pdfs, optimal maximum likelihood estimators of these
parameters can be obtained.

We have verified with WMAP-like simulations that the maximum likelihood estimation of the
\emph{quadratic} non-linear coupling parameters ($\hat{\rm f}_\nl$) is unbiased, at least
for a reasonable range of values, even when non-Gaussian simulations also account for a
\emph{cubic} term. In particular, we found that for simulated non-Gaussian coefficients
such as $|\bar{\rm f}_\nl| \lesssim 400$ and $|\bar{\rm g}_\nl| \lesssim 5\times 10^{5}$,
the estimation of f$_\nl$ is accurate and efficient. However, when trying to study
the case in which only a pure \emph{cubic} model is addressed, the situation is different.
In particular, the simulated \emph{quadratic} term has an important impact on the
estimation of g$_\nl$. For instance, if a pure \emph{quadratic} non-Gaussian model
is simulated with a value of $\bar{\rm f}_\nl \equiv 400$, a value of $\hat{\rm g}_\nl \approx 2.5\times 10^5$
is wrongly estimated. This results indicates, obviously, that the \emph{quadratic} term is
more important than the \emph{cubic} one in the expansion of the local non-Gaussian model, and, therefore, that
not accounting properly for the former might have a dramatic impact on the latter. Contrarily, the
opposite situation is much more unlikely.
Finally, we have investigated the joint estimation of the f$_\nl$ and g$_\nl$ parameters.
In this case we find that for a similar regime as the one mentioned above 
(i.e., $|\bar{\rm f}_\nl| \lesssim 400$ and $|\bar{\rm g}_\nl| \lesssim 5\times 10^{5}$),
an accurate and efficient estimation of the non-linear coupling parameters is obtained.
However, for larger values of these coefficients, we find that the parameter estimation is
highly biased, favouring a region of the parameter space of larger values for both
coefficients.

We have addressed, afterwards, the analysis of the WMAP 5-year data. We have consider two
different analyses. First, we have investigated the case of an all-sky analysis (except
for the Galactic area not allowed by the WMAP KQ75 mask). Second, and motivated by
previous findings, we have performed a separated analysis of two hemispheres. In particular, the
hemispherical division associated to the dipolar pattern found by~\cite{hoftuft09}
was considered.

Regarding the all-sky analysis, we find, for the case in which a pure \emph{quadratic}
model is investigated, the same result that we already found in our previous
work~\citep{vielva09}. In particular, we determine that $-154 < {\rm f}_\nl < 94$ at 95\%.
Equivalently, for the case of a pure \emph{cubic} non-Gaussian model we establish
$-5.6\times 10^5 < {\rm g}_\nl < 6.4\times 10^5$ at 95\%. This is in agreement  with a recent
work by~\cite{desjacques09}, where an analysis of LSS data was performed. The result that
we provide in this paper is, as far as we know, the first direct
constraint on g$_\nl$ from CMB data alone.
Finally, we have also investigated the case of a joint estimation of the \emph{quadratic}
and \emph{cubic} non-linear coupling parameters. In this case, the constraints obtained
on f$_\nl$ and g$_\nl$ are virtually the same as the ones already reported for the independent
analyses.

We have performed a model selection to evaluate which of the four hypotheses (i.e.
Gaussianity, a pure \emph{quadratic} model, a pure \emph{cubic} model and a
general non-Gaussian scenario) is more likely. Since a well motivated prior for the
non-linear coupling parameters is lacking, we have used asymptotic model selection
criteria (like AIC and BIC), instead of more powerful Bayesian approaches, like
the Bayesian Evidence or the posterior ratio test. Both methodologies (AIC and BIC)
indicates that the Gaussian hypothesis is more likely than any of the non-Gaussian
models. We also found that, among the non-Gaussian scenarios, the one with a pure \emph{quadratic}
model is the most favoured one, whereas the general one (i.e. f$_\nl \ne 0$ and g$_\nl \ne 0$)
is the most unlikely.

The analysis of the WMAP data in two hemispheres revealed that, whereas
both halves of the sky present similar constraints on the g$_\nl$ parameter (and, in
both cases, not indicating any significant incompatibility with the zero value), the
analysis of a pure \emph{quadratic} scenario showed a clear asymmetry. First, the f$_\nl$ value
in the hemisphere whose pole is closer to the Southern Ecliptic Pole, is
$\hat{\rm f}_\nl = -164 \pm 62$. Which implies that f$_\nl < 0$ at 96\%.
Even more, the distance between both likelihoods (given in terms of the KLD) presents a p-value
of $\approx 0.04$.

We have also analyzed the WMAP data after by considering different correlation properties in each
hemisphere (according to the dipolar modulation described by~\cite{hoftuft09}).
We have tested that the
behaviour found for the f$_\nl$ is practically the same as before, and that, in addition,
an extra anomaly appears associated with the g$_\nl$ parameter. In particular,
the distance between both likelihoods is anomalously large as well (it corresponds to
a p-value of $\lesssim 0.002$.
Further test was performed after correcting WMAP data from the dipolar modulation. In this case the
asymmetries in the maximum-likelihood estimations of both non-linear coupling parameters remain unaltered.
Hence, these results indicate that, as it has been previously reported
in other works, there are evidences of anisotropy in the WMAP data, reflected as
an asymmetry between two opposite hemispheres. Such anomaly is related to
a different distribution for a non-linear coupling parameter related to the
\emph{quadratic} term. However, a correction in
terms of a dipolar modulation as the one proposed by~\cite{hoftuft09}, seems
not to be sufficient to account for this anomaly related to the likelihood
distribution of the f$_\nl$ parameter.

\section*{Acknowledgements}
We thank Bel\'en Barreiro and Enrique Mart{\'\i}nez-Gonz\'alez for useful comments. We acknowledge partial financial support
from the Spanish Ministerio de Ciencia e Innovaci{\'o}n project
AYA2007-68058-C03-02. PV also thanks financial support from
the \emph{Ram\'on y Cajal} programme. The authors acknowledge the computer resources, technical
expertise and assistance provided by the Spanish Supercomputing
Network (RES) node at Universidad de Cantabria. We acknowledge the
use of Legacy Archive for Microwave Background Data Analysis
(LAMBDA). Support for it is provided by the NASA Office of Space
Science. The HEALPix package was used throughout the data analysis~\citep{gorski05}.

\appendix
\section{N-pdf derivation}
\label{sec:app}
In this Appendix we provide a detailed derivation of the N-pdf for the
non-linear local model given by equation~\ref{eq:model}. 
As it has been already discussed, 
the objective is
to make use of this likelihood to constrain the parameters that define the
perturbative model. In particular, we are interested in estimating the
parameters governing the \emph{quadratic} ($a$) and \emph{cubic} ($b$) terms,
that, as it was explained in Section~\ref{sec:model}, can be related (under
certain circumstances) to the non-linear coupling inflationary parameters f$_\nl$ and g$_\nl$, respectively.

Let us recall here the expression for the N-pdf for the non-Gaussian model,
as a function of the multinormal N-pdf ($p\left(\bmath{\phi} \right)$, equation~\ref{eq:pdfx}):
\begin{equation}
\label{eq:Apdfx} p\left(\bmath{x}\vert a,b \right) = p\left(\bmath{\phi} =
\bmath{\phi} \left(\bmath{x}\right)\right) Z,
\end{equation}
where $Z$ is the determinant of the Jacobian of the $\bmath{\phi} \longrightarrow \bmath{x}$ transformation (equation~\ref{eq:jacobian}). 
For practical purposes (i.e., in order to constrain the non-linear coupling parameters)
it is more convenient to deal with the log-likelihood ${\cal L}\left(\bmath{x}\vert a,b\right)$ (given in equation~\ref{eq:loglike}),
instead of $p\left(\bmath{x}\vert a,b\right)$ (given in equation~\ref{eq:pdfx}). Obviously, the later can
be easily obtained by the inversion of equation~\ref{eq:loglike}.

Replacing equation~\ref{eq:pdfx} into equation~\ref{eq:loglike}, it is easy to show that
the computation of ${\cal L}\left(\bmath{x}\vert a,b\right)$ implies to solve two terms:
\begin{equation}
\label{eq:partes}
{\cal L}\left(x_i\vert a,b\right) = \log Z + \Omega,
\end{equation}
where $\Omega \equiv \log{\frac{p\left( \bmath{\phi} = \bmath{\phi} \left(\bmath{x}\right) \right)}{p\left(\bmath{x} | 0\right)}}$,
and, we recall it, $p\left(\bmath{x} | 0\right) \equiv p\left(\bmath{\phi = x}\right)$.

Let us address the determination of these two terms ($\log Z$ and $\Omega$) independently.

\subsection{Determination of the log-Jacobian}
According to equation~\ref{eq:jacobian}, the log-Jacobian for the $\bmath{\phi} \longrightarrow \bmath{x}$ transformation is given by:
\begin{equation}
\label{eq:log-jacobian} 
\log Z = \sum_{i=1}^N \log \left( \frac{\partial \phi_i}{\partial x_i}\right), 
\end{equation}
where $\phi_i$ as a function of $x_i$ is given in equation~\ref{eq:inversmodel}.
It is straightforward to prove that:
\begin{equation}
\label{eq:partial} 
\frac{\partial \phi_i}{\partial x_i} = 1 + \sum_{m=1}^4 g_{m,i}\sigma^m,
\end{equation}
where
\begin{eqnarray}
\label{eq:gs} 
g_{1,i} & = & -2ax_i\nonumber \\
g_{2,i} & = & 3\left( 2a^2 - b \right)x_i^2 - 2a^2 \nonumber \\
g_{3,i} & = & 20\left( -a^3  + ab \right)x_i^3 + 6\left( 2a^3 - ab \right)x_i \nonumber \\
g_{4,i} & = & 5\left( 14a^4 -21a^2b + 3b^2\right)x_i^4 + 60(-a^4 + a^2b)x_i^2 \nonumber \\
& & + 3\left( 2a^4 - a^2b\right).
\end{eqnarray}
Since we are considering a perturbative non-Gaussian model, then $\sum_{m=1}^4 g_{m,i}\sigma^m \ll 1$ and, therefore, 
the log-Jacobian can be easily derived from the Taylor expansion for the logarithm function:
\begin{eqnarray}
\label{eq:logZexpan} 
\log Z & \approx & \sum_{i=1}^N \left( g_{1,i} \right)\sigma + \left( g_{2,i} - \frac{1}{2} g_{1,i}^2\right)\sigma^2  \\
& & + \left(g_{3,i} + \frac{1}{3}g_{1,i}^3 - g_{1,i}g_{2,i} \right)\sigma^3 \nonumber \\
& & + \left(g_{4,i} - \frac{1}{2}g_{2,i}^2 + g_{1,i}^2g_{2,1} - g_{1,i}g_{3,1} - \frac{1}{4}g_{1,i}^4 \right)\sigma^4 . \nonumber
\end{eqnarray}
Taking into account that: 
\begin{eqnarray}
\label{eq:xproperties} 
\frac{1}{N}\sum_{i=1}^N x_i^2 & = & 1 + 2\left( a^2 + 3b\right)\sigma^2 + {\cal O}\left(\sigma^3\right) \nonumber \\
\frac{1}{N}\sum_{i=1}^N x_i^3 & = & 6a\sigma + {\cal O}\left(\sigma^3\right) \nonumber \\
\frac{1}{N}\sum_{i=1}^N x_i^4 & = & 3 + {\cal O}\left(\sigma^2\right),
\end{eqnarray}
(as it can be trivially showed from equations~\ref{eq:model} and~\ref{eq:phiproperties})
and replacing equation~\ref{eq:gs} into equation~\ref{eq:logZexpan}, one finally gets:
\begin{equation}
\label{eq:finallogZ} 
\frac{1}{N}\log Z \approx \left(2a^2 - 3b\right)\sigma^2 + \left(12a^4 - 36a^2b + \frac{27}{2} \right)\sigma^4,
\end{equation}
up to the appropriate order in $\sigma$.

\subsection{Determination of $\Omega$}
Since we are dealing with a perturbative model (i.e., $\bmath{\phi} = \bmath{x} + \bmath{\epsilon}$, with $\bmath{\epsilon} \ll \bmath{x}$
and where $\bmath{\epsilon} \equiv \bmath{\eta}\sigma + \bmath{\nu}\sigma^2 + \bmath{\mu}\sigma^3 + \bmath{\lambda}\sigma^4$,
according to equation~\ref{eq:inversmodel})
it is interesting to notice that the probability function
$p\left(\bmath{\phi} = \bmath{\phi} \left(\bmath{x}\right)\right)$ can be expanded in terms of the Taylor
expansion, up to the appropriate order:
\begin{eqnarray}
\label{eq:pdf_expasion} 
p\left(\bmath{\phi} = \bmath{\phi} \left(\bmath{x}\right) \right) & = &  P + \epsilon^i \partial_i P + \frac{1}{2}\epsilon^i \epsilon^j \partial_{ij} P
+ \frac{1}{6}\epsilon^i \epsilon^j \epsilon^k \partial_{ijk} P \nonumber \\
& &  + \frac{1}{24}\epsilon^i \epsilon^j \epsilon^k \epsilon^l \partial_{ijkl} P + {\cal O}\left( 5 \right),
\end{eqnarray}
where, for simplicity, we define $P \equiv p\left( \bmath{\phi} = \bmath{x} + 0 \right) \equiv p\left( \bmath{x} | 0 \right)$. 
The function $\partial_i Y\left(\bmath{x}\right)$ is the derivative of
$Y\left(\bmath{x}\right)$ with respect to $x_i$, i.e.: $\partial_i Y\left(\bmath{x}\right) \equiv \frac{\partial Y\left(\bmath{x}\right)}{\partial x_i}$.
Equivalently, $\partial_{ij} Y\left(\bmath{x}\right) \equiv \frac{\partial}{\partial x_j} \partial_i Y\left(\bmath{x}\right)$, 
$\partial_{ijk} Y\left(\bmath{x}\right) \equiv \frac{\partial}{\partial x_k} \partial_{ij} Y\left(\bmath{x}\right)$,
and 
$\partial_{ijkl} Y\left(\bmath{x}\right) \equiv \frac{\partial}{\partial x_l}  \partial_{ijk} Y\left(\bmath{x}\right)$.

Taking into account this expansion, we have::
\begin{eqnarray}
\label{eq:omega1} 
\frac{p\left(\bmath{\phi} = \bmath{\phi} \left(\bmath{x}\right) \right)}{p\left( \bmath{x} | 0 \right)} & = &  1 + A^i \epsilon_i + \frac{1}{2} A^{ij}\epsilon^i \epsilon^j + \frac{1}{6}A^{ijk} \epsilon^i \epsilon^j \epsilon^k  \nonumber \\
& &  + \frac{1}{24} A^{ijkl} \epsilon^i \epsilon^j \epsilon^k \epsilon^l,
\end{eqnarray}
where, $A^{i} \equiv \frac{\partial_{i}P}{P}$, $A^{ij} \equiv \frac{\partial_{ij}P}{P}$, $A^{ijk} \equiv \frac{\partial_{ijk}P}{P}$,
and $A^{ijkl} \equiv \frac{\partial_{ijkl}P}{P}$. It is trivial to show that these quantities are related to the data $\bmath{x}$,
its expected correlation $\bmath{\xi} = \left[ \xi_{ij} \right]$, and the model (i.e., $a$ and $b$). After some algebra, one trivially finds:
\begin{eqnarray}
\label{eq:as} 
A_{i} & = & -\xij x^{j},\\
A_{ij} & = & -\xij +  A_{i}A_{j},\nonumber \\
A_{ijk} & = & A_{ij}A_{k} - A_{j}\xik - A_{i}\xjk \nonumber \\
A_{ijkl} & = & A_{i}A_{j}A_{k}A_{l} + \xij\xkl + \xik\xjl + \xjk\xil \nonumber \\
& & -A_{k}A_{l}\xij -A_{j}A_{l}\xik -A_{i}A_{l}\xjk \nonumber \\
& & -A_{k}A_{i}\xjl -A_{k}A_{j}\xil -A_{i}A_{j}\xkl \nonumber.
\end{eqnarray}

Since we are interested in the quantity $\log{\frac{p\left( \bmath{\phi} = \bmath{\phi} \left(\bmath{x}\right) \right)}{p\left(\bmath{x} | 0\right)}}$, 
the logarithm function can be expanded up to the appropriate order. It is straightforward to obtain:
\begin{equation}
\label{eq:omega2} 
\Omega = F\sigma + G\sigma^2 + H\sigma^3 + I\sigma^4 + {\cal O}\left( 5 \right),
\end{equation}
where:
\begin{eqnarray}
\label{eq:coeffs} 
F & = & -\frac{1}{N} \left( S_{\bmath{\eta}\bmath{x}} \right),\\
G & = & -\frac{1}{N} \left( S_{\bmath{\nu}\bmath{x}} + \frac{1}{2}S_{\bmath{\eta}\bmath{\eta}} \right), \nonumber\\
H & = & -\frac{1}{N} \left( S_{\bmath{\lambda}\bmath{x}} + S_{\bmath{\eta}\bmath{\nu}} \right), \nonumber\\
I & = & -\frac{1}{N} \left( S_{\bmath{\mu}\bmath{x}} + S_{\bmath{\lambda}\bmath{\eta}} + \frac{1}{2}S_{\bmath{\nu}\bmath{\nu}} \right). \nonumber
\end{eqnarray}
We have defined the operator $S_{\bmath{\alpha}\bmath{\beta}} \equiv \alpha^{i} \xij \beta^{j}$, and $\bmath{\eta}$, $\bmath{\nu}$,
$\bmath{\lambda}$ and $\bmath{\mu}$ where defined in equation~\ref{eq:numu}.
\\
\\

Therefore, taking into account equations~\ref{eq:finallogZ} and~\ref{eq:omega2}, the final
expression for the log-likelihood ${\cal L}\left(\bmath{x}\vert a,b\right)$ is:
\begin{eqnarray}
\label{eq:loglikefinal} 
\frac{1}{N} {\cal L}\left(\bmath{x}\vert a,b\right) & = & F\sigma + \left(2a^2 - 3b + G\right)\sigma^2  + H\sigma^3 \nonumber \\
& & + \left(12a^4 - 36a^2b + \frac{27}{2}b^2 + I \right)\sigma^4.
\end{eqnarray}

\label{lastpage}

\end{document}